\documentclass[twocolumn,english,prb,notitlepage,superscriptaddress]{revtex4-2}
\usepackage[T1]{fontenc}
\usepackage[latin9]{inputenc}
\usepackage{graphicx}
\usepackage{amsmath}
\usepackage{float}
\usepackage[colorlinks = true,
linkcolor = blue,
urlcolor  = blue,
citecolor = blue,
anchorcolor = blue]{hyperref}
\usepackage[usenames, dvipsnames]{color}

\makeatletter
\usepackage{babel}
\usepackage{float}

\usepackage{tabularx}
\newcolumntype{C}{>{\centering\arraybackslash}X}

\makeatother

\begin{document}
	
%	%%% This disclaimer is mandatory for all ORNL preprints
%	\setcounter{page}{0}
%	Notice: This manuscript has been authored by UT-Batelle, LLC, under contract DE-AC05-00OR22725 with the US Department of Energy (DOE). The US government retains and the publisher, by accepting the article for publication, acknowledges that the US government retains a nonexclusive, paid-up, irrevocable, worldwide license to publish or reproduce the published form of this manuscript, or allow others to do so, for US government purposes. DOE will provide public access to these results of federally sponsored research in accordance with the DOE Public Access Plan (http://energy.gov/downloads/doe-public-access-plan).
%	
%	\newpage
%	
%	\quad
%	
%	\newpage

\title{Spin waves and magnetic exchange Hamiltonian in CrSBr}

\author{A. Scheie}  % Approved 
\email{scheieao@ornl.gov}
\address{Neutron Scattering Division, Oak Ridge National Laboratory, Oak Ridge, Tennessee 37831, USA}

\author{M. Ziebel} % Approved 
\address{Department of Chemistry, Columbia University, New York, NY 10027, USA}

\author{D. G. Chica} % Approved (4/18/22)
\address{Department of Chemistry, Columbia University, New York, NY 10027, USA}

\author{Y. J. Bae}  % Approved (4/14/22)
\address{Department of Chemistry, Columbia University, New York, NY 10027, USA}

\author{Xiaoping Wang}  % Approved (4/13/22)
\address{Neutron Scattering Division, Oak Ridge National Laboratory, Oak Ridge, Tennessee 37831, USA}

\author{A. I. Kolesnikov}  % Approved (4/11/22)
\address{Neutron Scattering Division, Oak Ridge National Laboratory, Oak Ridge, Tennessee 37831, USA}

\author{Xiaoyang Zhu}
\address{Department of Chemistry, Columbia University, New York, NY 10027, USA}

\author{X. Roy}
\address{Department of Chemistry, Columbia University, New York, NY 10027, USA}

\date{\today}

%Note: abstract needs to be under 600
\begin{abstract}
% Draft figures for CrSBr spin wave paper. Figures \ref{fig:crystalstruct} - \ref{fig:SpinWavesDM} are for the main text, and Figures \ref{fig:sample} - \ref{fig:Chisq_vs_n_DM} are for the appendices.

CrSBr is an air-stable 2D van der Waals semiconducting magnet with great technological promise, but its atomic-scale magnetic interactions---crucial information for high-frequency switching---are poorly understood. We present an experimental study to determine the CrSBr magnetic exchange Hamiltonian and bulk magnon spectrum. We confirm the $A$-type antiferromagnetic order using single crystal neutron diffraction. We also measure the magnon dispersions using inelastic neutron scattering and rigorously fit the excitation modes to a spin wave model. The magnon spectrum is well described by an intra-plane ferromagnetic Heisenberg exchange model with seven nearest in-plane exchanges. %; anisotropic exchange and inter-layer exchange are too weak to resolve. 
This fitted exchange Hamiltonian enables theoretical predictions of CrSBr behavior: as one example, we use the fitted Hamiltonian to predict the presence of chiral magnon edge modes with a spin-orbit enhanced CrSBr heterostructure.

\end{abstract}

\maketitle

\section{Introduction}

Two-dimensional (2D) magnetism has long been a topic of theoretical investigation, but only recently has it become experimentally accessible through van der Waals materials \cite{Huang2017,gong2017discovery}.
With monolayer magnetism preserved through magnetic anisotropy, these materials promise to yield cleaner experimental realizations of theoretical states, novel spintronic devices, and new topological phases of matter \cite{Burch2018,Chen_2018}.
However, accurately predicting the magnetic properties and excitations %, especially for transport at short timescales relevant for electronic switching, 
requires a detailed knowledge of the magnetic exchange Hamiltonian.

\begin{figure}
	\centering\includegraphics[width=0.46\textwidth]{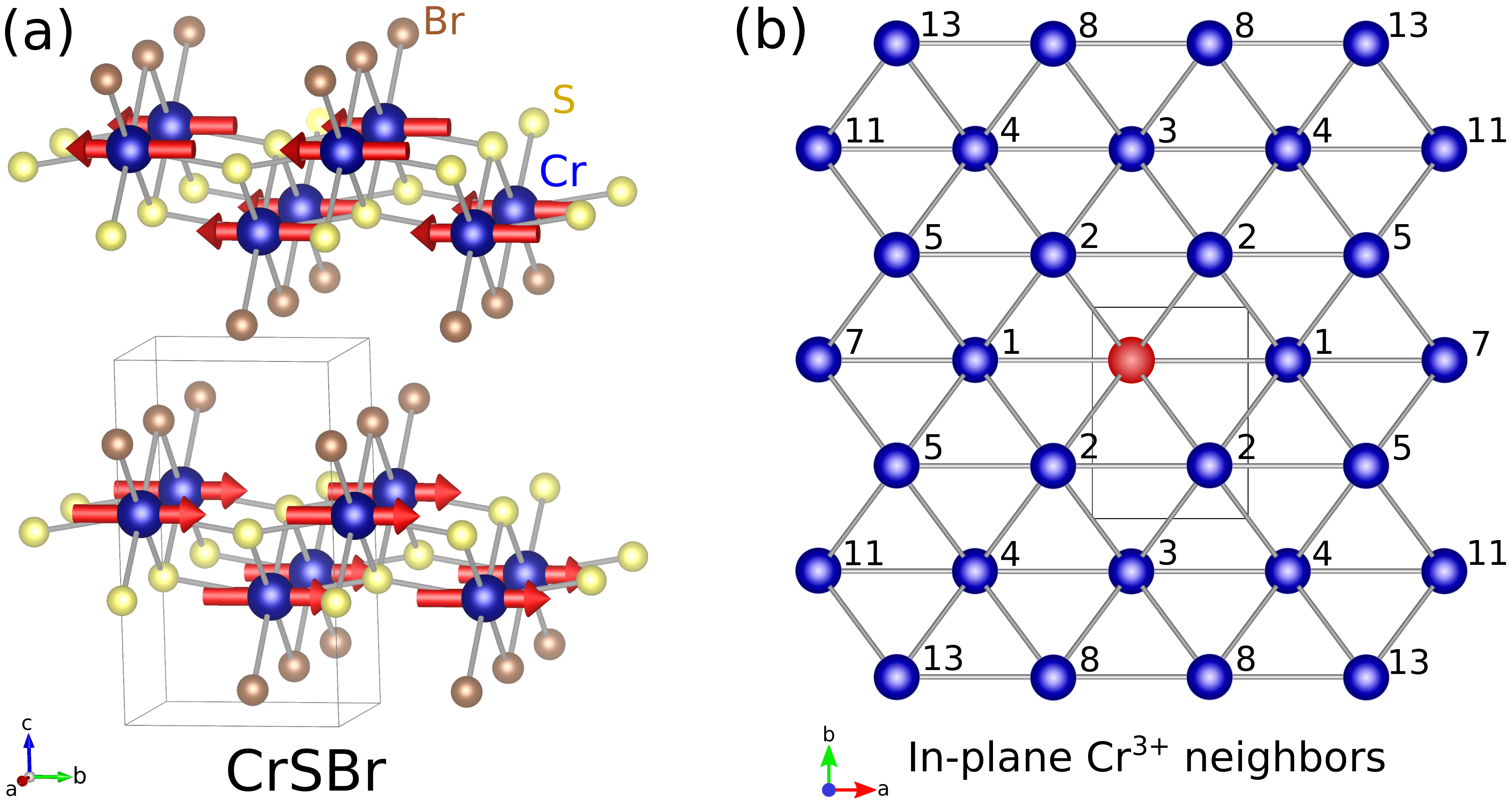}
	\caption{CrSBr crystal structure. (a) shows the crystal structure and Cr$^{3+}$ magnetic order, ferromagnetic in-plane but layered in alternating directions for a bulk antiferromagnetism. (b) shows the Cr neighbors in the plane from the central red atom, numbered in order of bond length (neighbors 6, 9, 10, and 12 are between planes).}
	\label{fig:crystalstruct}
\end{figure}

A promising 2D van der Waals magnet is CrSBr. CrSBr forms in 2D layers of magnetic Cr$^{3+}$ ions forming a rectangular lattice, as shown in Fig. \ref{fig:crystalstruct}. In bulk, it orders magnetically at $T_N=132$~K  \cite{goser1990magnetic,telford2020layered,lopez2022dynamic} with A-type antiferromagnetism: ferromagnetic planes polarized along the $b$ axis, alternating in orientation for an overall antiferromagnetic (AFM) order, and becoming ferromagnetic (FM) in the monolayer limit \cite{lee2021magnetic}. This material is air-stable and has a semiconducting gap low enough to gate with realistic electric fields \cite{qi2018electrically,telford2020layered}. There is also intricate interplay between electronic transport, optical properties, and magnetism \cite{telford2021hidden,Wilson2021}, with the potential for exploiting both spin and charge degrees of freedom for technological purposes \cite{Ghiasi2021}. 
To understand, predict, and ultimately exploit the spin transport properties of CrSBr, it is necessary to know the magnetic exchange Hamiltonian between Cr ions and the resulting magnon dispersions. In particular, the high-frequency behavior of the magnon bands is critical to understanding the short-time behavior relevant for electronic switching and information processing. In this study, we measure the static magnetic structure and the high energy magnon dispersions, experimentally determine the spin exchange Hamiltonian using inelastic neutron scattering, and then use this Hamiltonian to predict the presence of chiral edge modes in layered heterostructures.

\section{Results and Analysis}

\subsection{Static magnetism}

\begin{figure}
	\centering\includegraphics[width=0.49\textwidth]{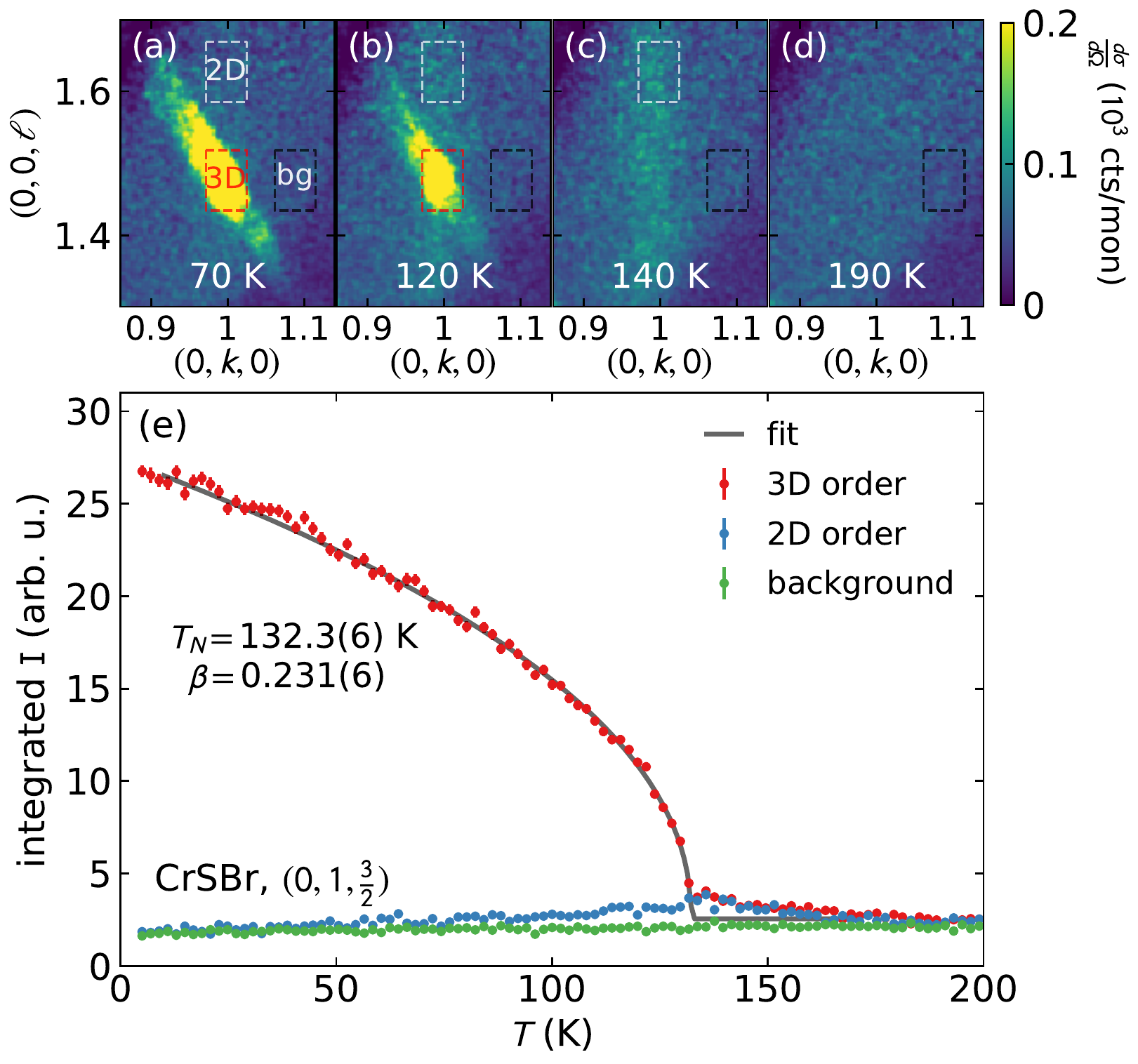}
	\caption{Single crystal CrSBr neutron diffraction. (a)-(d) show the temperature evolution of the $(0,1,\frac{3}{2})$ Bragg peak (which is smeared out in $Q_{\perp}$ due to crystal twinning). Above and near the phase transition, a streak of scattering along $\ell$ appears, signaling two-dimensional (2D) magnetic correlations. We track the 3D and 2D correlations using the red and white boxes in panels (a)-(d) (the black box is the background), plotted in panel (e). The 2D correlations peak at the ordering temperature $T_N=132.2(6)$~K, and decrease at lower temperatures. The fitted order parameter curve is shown in grey, with a fitted $\beta=0.231(6)$.}
	\label{fig:elastic}
\end{figure}

The single crystal neutron diffraction is shown in Fig. \ref{fig:elastic} and confirms the ground state magnetic order in Ref. \cite{lopez2022dynamic}: below a transition temperature of 132.3(6)~K, new Bragg peaks appear at half-integer $\ell$ positions in accord with $(00\frac{1}{2})$ magnetic order. In the Supplemental Information, we refine the Bragg intensities and show they indicate A-type antiferromagnetism in Fig. \ref{fig:crystalstruct}. At temperatures near $T_N$, a streak of scattering appears at $(0,1,\ell)$, signaling 2D magnetic correlations in the $ab$ plane. Tracking the 2D correlations as a function of temperature, we see they peak at $T_N$, but with significant 2D magnetic correlations above $T_N$. %, which is consistent with the monolayer CrSBr having a higher magnetic ordering temperature than the bulk \cite{lee2021magnetic}.

The 3D Bragg intensity vs temperature follows a smooth curve between $T_N$ and 5 K. Although CrSBr samples show a sample-dependent discontinuity in magnetic susceptibility at 30~K \cite{telford2021hidden}, no such feature is observed in the neutron diffraction. Furthermore, aside from a larger static moment at 5~K, we also find no difference between 80~K and 5~K refined magnetic structures. Thus we conclude, as did Ref. \cite{lopez2022dynamic}, that the 30~K discontinuity is not associated with a change in the spatially-averaged magnetic order. This is consistent with the proposal in Ref. \cite{telford2021hidden} that the susceptibility discontinuity is due to local or impurity spins. 
Fitting the 3D Bragg intensity to an order parameter curve, we find a critical exponent $\beta=0.231\pm0.006$. This is far from the theoretical 3D Heisenberg critical exponent $\beta = 0.36$ \cite{chaikin1995principles}. Instead, this is remarkably close to the critical exponent $\beta=0.231$ derived for the 2D $XY$ model via Kosterlitz-Thouless (K-T) theory \cite{Bramwell_1993}, showing very 2D exchange interactions with easy-plane anisotropy, in accord with expected Van der Waals behavior. 

\subsection{Dynamic magnetism}

\begin{figure*}
	\centering\includegraphics[width=0.99\textwidth]{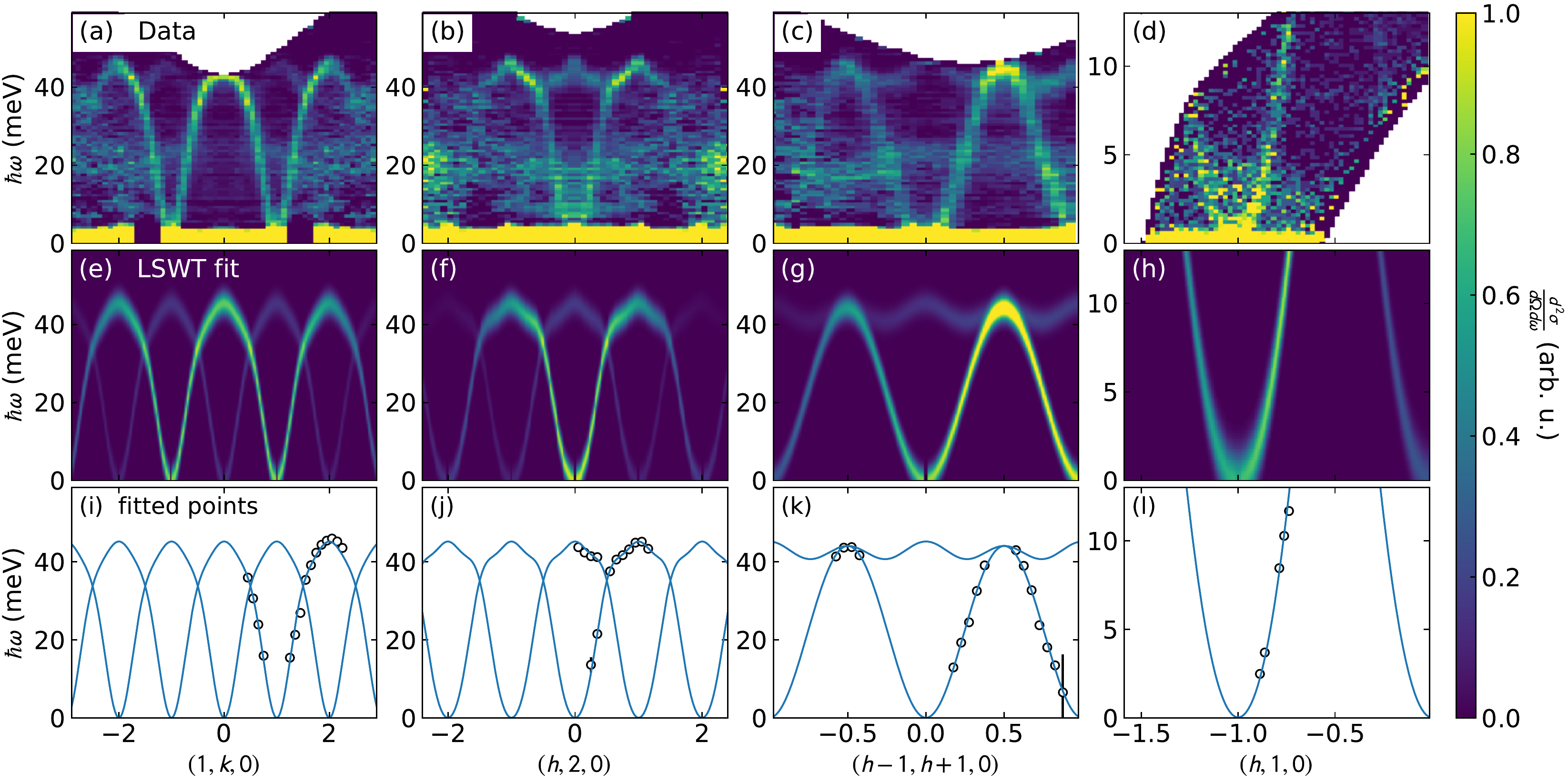}
	\caption{Measured and fitted spin wave spectra of CrSBr. The top row (a)-(d) shows the measured spin wave spectra of CrSBr. Panels (a)-(c) were measured with $E_i=70$~meV neutrons, while panel (d) was measured with $E_i=20$~meV neutrons. The middle row (e)-(h) shows the LSWT calculated spectrum from the best fit Hamiltonian in Table \ref{tab:FittedJ}. The bottom row (i)-(l) shows a portion of the data points used in the fit (black circles), and the fitted dispersion (blue solid line). For a complete list of fitted data, see the Supplemental Information.}
	\label{fig:SpinWaves}
\end{figure*}

Several plots of CrSBr inelastic neutron scattering data are shown in Fig. \ref{fig:SpinWaves}. Because of the small sample mass, there is substantial background noise from phonon scattering in the aluminum sample holder. Nevertheless, the magnon modes are clearly distinguished by (i) their symmetries following the CrSBr reciprocal lattice units, (ii) their intensities following a magnetic form factor with intensity largest near $|Q|=0$, and (iii) comparison with a measured background (see section \ref{sec:Experimental}).

 The magnon dispersions in the $(hk0)$ plane reach a maximum energy of $\sim 45$~meV. To within an energy resolution of $\pm 0.5$~meV FWHM in the $E_i = 20$~meV data in Fig. \ref{fig:SpinWaves}(d), the modes are gapless at $h+k=$ even integer points in reciprocal lattice units (RLU). This lack of observable gap evidences highly isotropic magnetism, as one expects for $S=3/2$ Cr$^{3+}$. This comports with density functional theory predictions \cite{Wang_2020_Electrically} and recent photon measurements finding $Q=0$ magnon gaps of 0.102(3)~meV and 0.141(4)~meV \cite{bae2022exciton}, as well as magnetization measurements finding a maximum anisotropy of 0.144~$\rm \mu eV$ at 2~K ($c$-axis compared to $b$) in CrSBr (see Supplemental Information \cite{SuppMat}): too small to be resolved in this experiment.

\begin{figure}
	\centering\includegraphics[width=0.48\textwidth]{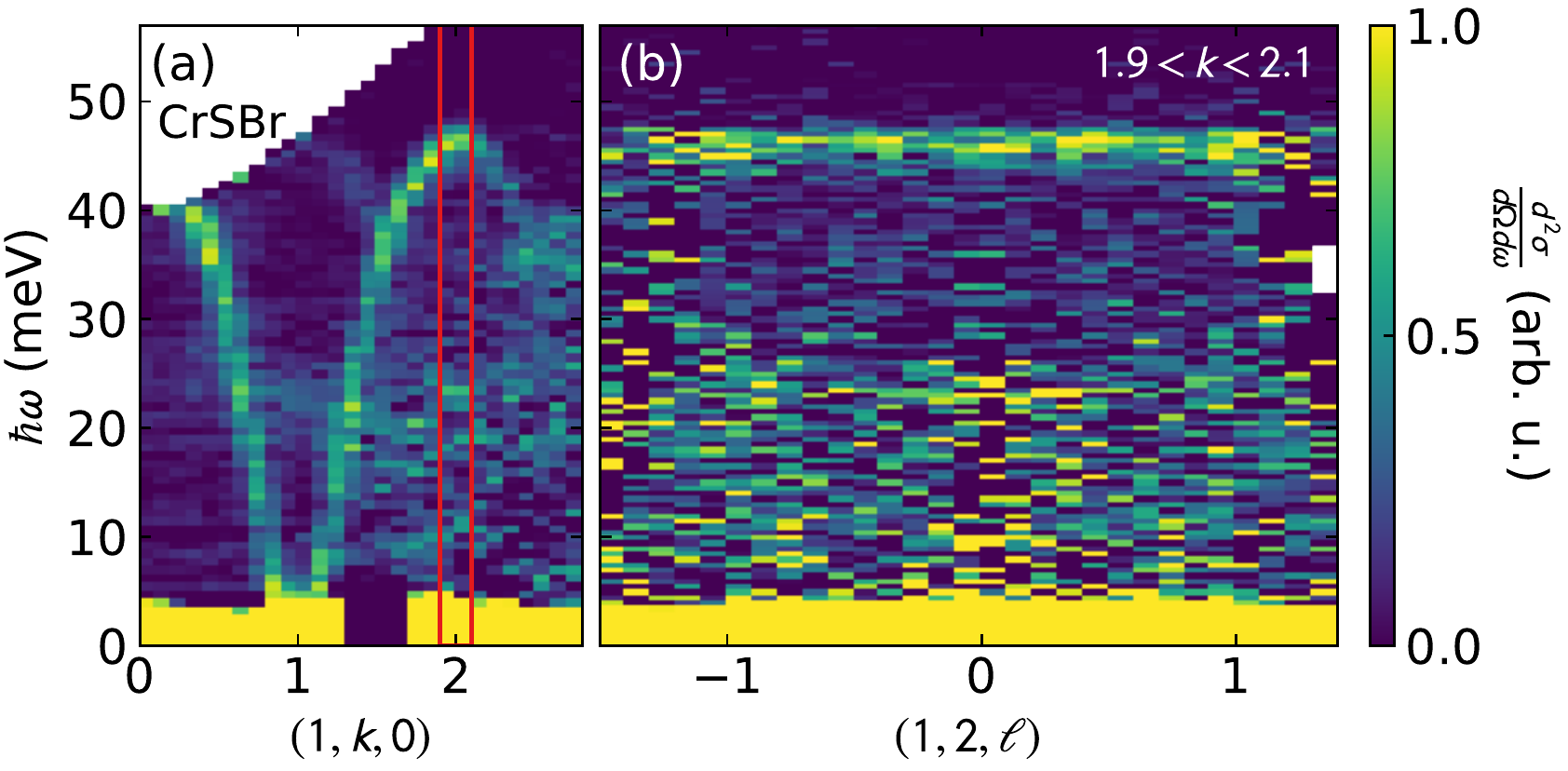}
	\caption{CrSBr dispersion along $\ell$. Panel (a) shows a cut along $(1,k,0)$ with red lines delineating the cut in panel (b) along $(1,2,\ell)$. There is no detectable dispersion along $\ell$ at this or any other wavevector, showing that the inter-plane magnetic exchange is negligibly weak.}
	\label{fig:Ldependence}
\end{figure}

In the $\ell$ direction, we find no measurable dispersion out of plane at all $h$ and $k$, as shown in Fig. \ref{fig:Ldependence}. This evidences very weak inter-plane magnetic exchange, as one would expect for a highly 2D system (see the Supplemental Information for further details \cite{SuppMat}). This is consistent with the photon excitation study in Ref \cite{bae2022exciton} which finds an interlayer exchange $< 0.01$~meV, as well as density functional calculations in Ref. \cite{Yang_2021} which finds a CrSBr interlayer magnetic interaction three orders of magnitude weaker than the in-plane interactions. 
Because the modes are flat with $\ell$, all in-plane scattering data presented here is integrated over $-1<\ell<1$ RLU to maximize the magnon mode visibility.

\subsection{Fitting the exchange Hamiltonian}

We determined the CrSBr magnetic exchange constants from this scattering data by performing a fit to a linear spin wave theory (LSWT) model. 
The spin wave model for a bipartite ferromagnetic lattice is calculated following Ref. \cite{Jensen+Mackintosh} using the Hamiltonian
\begin{equation}
\mathcal{H} = \sum_{i,j} J_{\langle ij \rangle} \vec{S_i} \cdot \vec{S_j} \label{eq:Hamiltonian}
\end{equation}
where $\vec{S_i}$ are vectors of length $|\vec{S_i}|=3/2$ and  $J_{\langle ij \rangle}$ are magnetic exchange constants between pairs of spins. Because many exchanges are symmetry-equivalent, we write  $J_n$ where $n$ is the neighbor number. The fitted neighbors $n$ are shown in Fig. \ref{fig:crystalstruct}(b).

To constrain the fit, we extracted 188 unique $Q$ and $\hbar \omega$ points by fitting constant $|Q|$ cuts of the magnon modes to Gaussian profiles in energy across 11 different data slices, using only regions where the magnons are clearly distinguishable from background (see Supplemental information for details \cite{SuppMat}). We then defined a global reduced $\chi^2$ function based on magnon mode energies at those $Q$ points, minimizing $\chi^2_{red}$ by varying $J_n$  using Scipy's optimization package \cite{virtanen2020scipy}.

\begin{figure}
	\centering\includegraphics[width=0.49\textwidth]{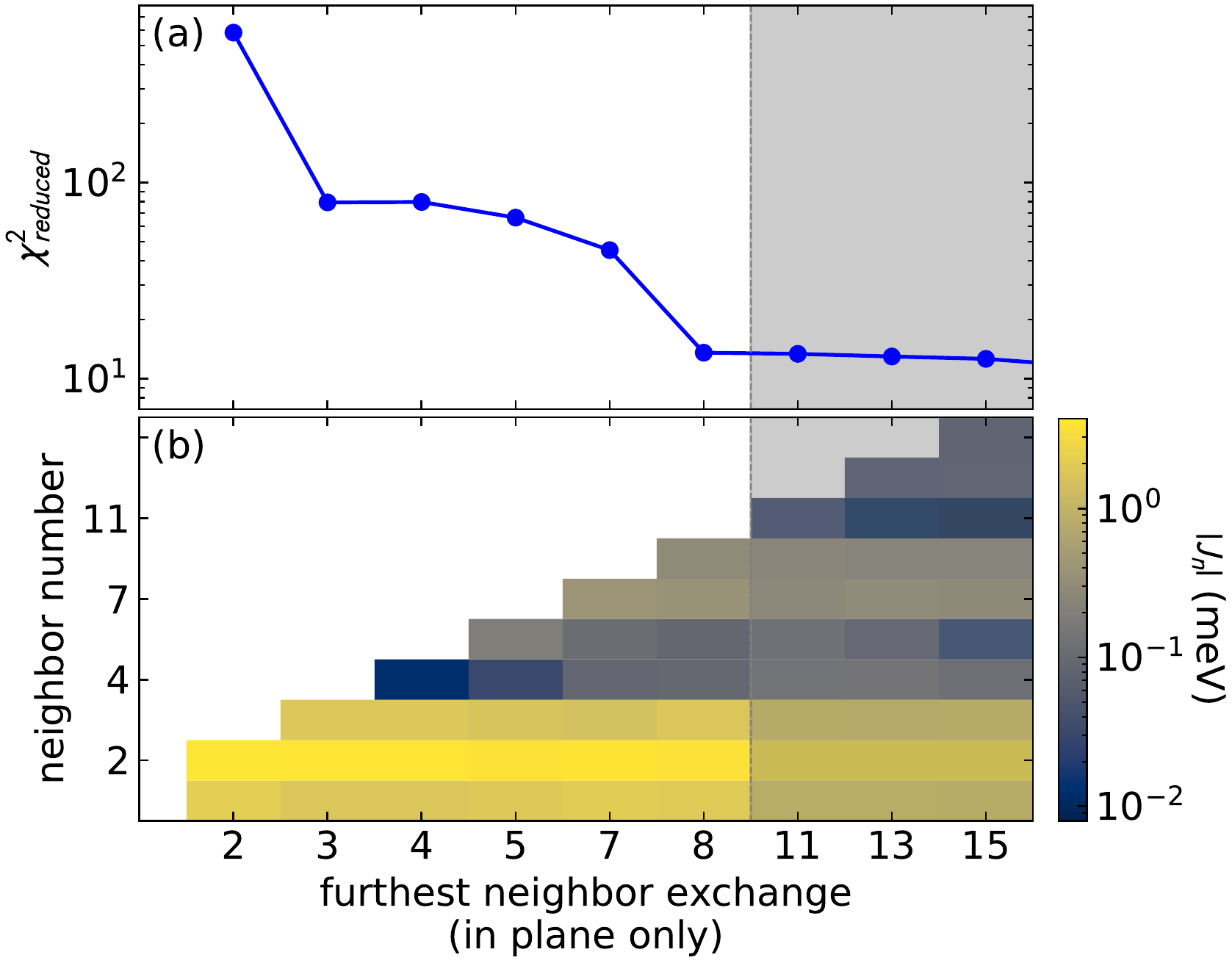}
	\caption{Dependence of the best fit $\chi_{\rm red}^2$ on the number of neighbors included in the fit. Panel (a) shows $\chi_{\rm red}^2$ vs neighbor number $n$, and panel (b) shows these fitted values in a colormap. Beyond the 8th neighbor, the $\chi_{\rm red}^2$ does not appreciably improve by adding additional neighbors, so we truncate our model at the 8th neighbor.}
	\label{fig:Chisq_vs_n}
\end{figure}

To systematically determine the number of exchange constants to include in our model, we fitted the magnon modes to a spin wave model beginning with only two neighbors, and increasing the number of neighbors up to the 17th neighbor exchange  (excluding all inter-plane exchanges), re-fitting for each new neighbor. We find that additional neighbors improve the best fit $\chi^2_{red}$ value up to the 8th neighbor. Including neighbors beyond 8 does not improve $\chi^2_{red}$ by a significant amount, as shown in Fig. \ref{fig:Chisq_vs_n}. Furthermore, we find that the statistical uncertainty of all exchanges beyond the 8th neighbor overlap with zero, and so we truncate our model at the 8th neighbor exchange and consider all further exchanges to be negligible in CrSBr. 

The best fit CrSBr Hamiltonian is given in Table \ref{tab:FittedJ}.
Uncertainty was calculated via a $\Delta \chi^2_{red} = 1$ contour for a one standard deviation statistical uncertainty \cite{NumericalRecipes}, see Supplemental Information for details \cite{SuppMat}. This was added in quadrature to the systematic uncertainty from truncating the model to the 8th neighbor exchange, taken to be the range of parameter variation between $n=11$ and $n=17$ fits. 
We simulated the neutron cross section for this best fit Hamiltonian using \textit{SpinW} software package \cite{SpinW}, plotted in Fig. \ref{fig:SpinWaves}(e)-(h). % The agreement between experiment and theoretical simulation is excellent for all measured wavevectors.

\begin{table}
	\caption{Best fit Hamiltonian exchange parameters for CrSBr.}
	\begin{ruledtabular}
		\begin{tabular}{cccc}
$J_{1}=$  &  $ -1.90 \pm 0.10$ meV &  $J_{5}=$  &  $ -0.09 \pm 0.06$ meV \\
$J_{2}=$  &  $ -3.38 \pm 0.06$ meV &  $J_{7}=$  &  $ 0.37 \pm 0.09$ meV \\
$J_{3}=$  &  $ -1.67 \pm 0.10$ meV &  $J_{8}=$  &  $ -0.29 \pm 0.05$ meV \\
$J_{4}=$  &  $ -0.09 \pm 0.05$ meV &   \\
	\end{tabular}\end{ruledtabular}
\label{tab:FittedJ}
\end{table}

The agreement between theory and experiment is remarkably good for this isotropic exchange model. However, asymmetric Dzyaloshinskii-Moriya (DM) exchange  $\mathcal{H} = \vec{D}_{\langle ij \rangle} \cdot (\vec{S_i} \times \vec{S_j})$ is symmetry-allowed on the nearest neighbor Cr-Cr bond, with a $\vec{D}_{1}$ vector along the $b$ direction \cite{Moriya_1960}. This DM interaction would produce a magnon mode splitting at half-integer $k$ wavevectors. Although this exchange is expected to be weak in Cr$^{3+}$ because of its small spin-orbit coupling, such mode splitting was observed in CrI$_3$ with a fitted DM interaction of 0.31~meV \cite{Chen_2018}. Thus it may be that a weak DM exchange also plays a role in CrSBr.

\begin{figure}
	\centering\includegraphics[width=0.49\textwidth]{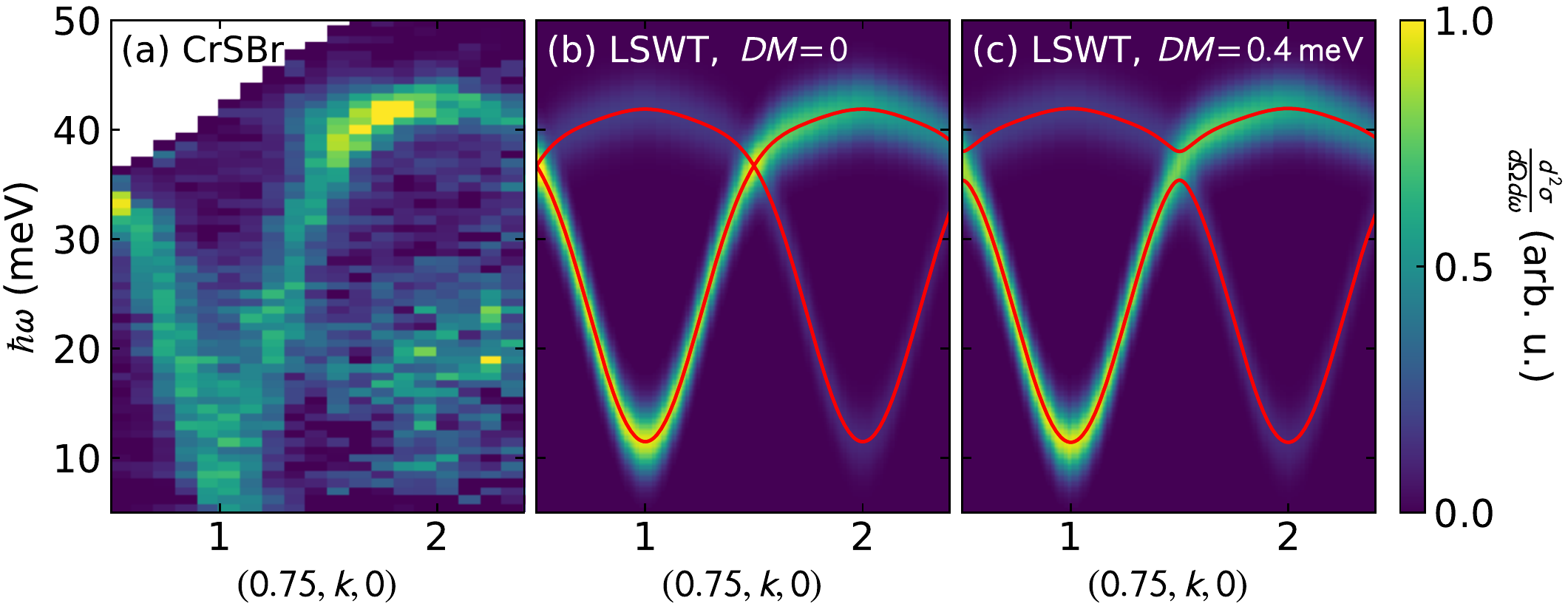}
	\caption{Effect of nearest neighbor DM interaction on the CrSBr dispersion. Panel (a) shows the CrSBr data along $(0.75,k,0)$, and panels (b) and (c) show the LSWT predictions with and without a DM term. The DM induces a gap at $k=1.5$, but no gap is resolvable in the data. This constrains the nearest neighbor DM term to be $< 0.8$~meV.}
	\label{fig:SpinWavesDM}
\end{figure}

To test whether the DM exchange is significant, we added a nearest neighbor DM exchange to our fitted model and allowed it to vary along with the other fitted parameters. No mode splitting is observed in our data, so any split modes are below the experimental resolution (see Fig. \ref{fig:SpinWavesDM}). We find that the best fit nearest neighbor $\vec{D}_1$ is unstable against the number of neighbors $n$ included in the model, varying between 0.0~meV and 0.4(4)~meV. We also find that the uncertainty overlaps with zero for all $n$. Furthermore, the best fit $\chi^2_{red}$ slightly \textit{worsens} when the DM exchange is added: $\chi^2_{red}=13.5819$ with $\vec{D}_1$, $\chi^2_{red}=13.5818$ without $\vec{D}_1$ (see Supplemental Information \cite{SuppMat}). Therefore, we consider the DM exchange to be negligible for CrSBr. While it is presumably nonzero, it is too small to resolve using this data.

\section{Discussion}

These results show that the CrSBr spin exchange Hamiltonian can be accurately approximated as a single-layer ferromagnet. Single-ion anisotropy, inter-plane exchange, and anisotropic exchange are all too small to resolve, leaving the exchange constants in Table \ref{tab:FittedJ} as an effective minimal model for the high frequency (short time) behavior of CrSBr. 
The fitted exchange parameters are almost uniformly ferromagnetic, with very similar exchange in the $a$ and $b$ directions, evidencing very two-dimensional magnetism (in contrast to the quasi-1D electronic bands \cite{Wu_2022_quasi1d}).
The CrSBr magnetic Hamiltonian having significant magnetic exchange out to the 8th neighbor is somewhat surprising, but is consistent with the strong Cr-S and Cr-Br covalency \cite{Yang_2021} which gives opportunity for extended orbital overlap.

We can compare this with first principles predictions for CrSBr. Guo et al \cite{guo2018chromium} used density functional theory to predict $J_1 = -1.72$~meV,   $J_2 = -3.25$~meV for CrSBr (normalized to the $S=3/2$ vector convention we use in Eq. \ref{eq:Hamiltonian}). This is very close to the fitted $J_1=-1.9(1)$~meV and $J_2=-3.38(6)$~meV, showing good agreement between experiment and theory.
Similarly, Wang et al \cite{Wang_2020_Electrically} and Yang et al \cite{Yang_2021} also used density functional theory on to predict weak CrSBr single-ion anisotropy (too weak to be measured with our measurements), although both their calculated CrSBr bulk exchange constants %($J_1=-2.79$~meV, $J_2=-3.88$~meV, $J_3=-2.12$~meV) 
are larger than we observe in experiment. 

Because the CrSBr semiconducting gap is 1.25(7)~eV \cite{telford2020layered} (14500~K), the effects of thermally populated conduction-mediated exchange will be very minor between 5~K and 300~K. Some exchange constant shifts with lattice expansion is possible, but such effects will also be minor \cite{Boix_2022_Probing}. Therefore we expect the magnetic exchange constants in Table \ref{tab:FittedJ} can be considered approximately correct at all temperatures below 300~K.

% Edge modes
\subsection*{Calculating edge modes}
Having determined the spin exchange Hamiltonian for CrSBr, we can begin using it to calculate relevant quantities. 
Among many spintronics proposals are ``topological magnonics'': using magnon edge modes for low-dissipation transport and switches \cite{Wang_2018_topological}. Magnon edge states, which only exist on the edge of a 2D material, generally have different dispersions than those in the bulk. For certain lattice geometries and Hamiltonians, the edge magnons can be ``chiral'', with a directional velocity preference based on the terminating surface \cite{Barman_2021}. Such chiral edge modes can be induced in ferromagnets via an anisotropic DM interaction \cite{McClarty_2021}.

In 2D materials, it is possible to increase the anisotropy via proximity effects with layers of heavy atoms, thereby enhancing spin-orbit interaction \cite{Bihlmayer_2015,Fan2014}. This has been powerfully demonstrated with graphene heterostructures \cite{avsar2014spin,Wang_2016_graphene}. Because spin orbit interaction drives the asymmetric DM exchange \cite{Moriya_1960}, it is possible to increase the CrSBr DM interaction via layering with a strong spin-orbit coupled material \cite{Yang_2020_Creation,wu2020neel}. 

\begin{figure}
	\centering\includegraphics[width=0.45\textwidth]{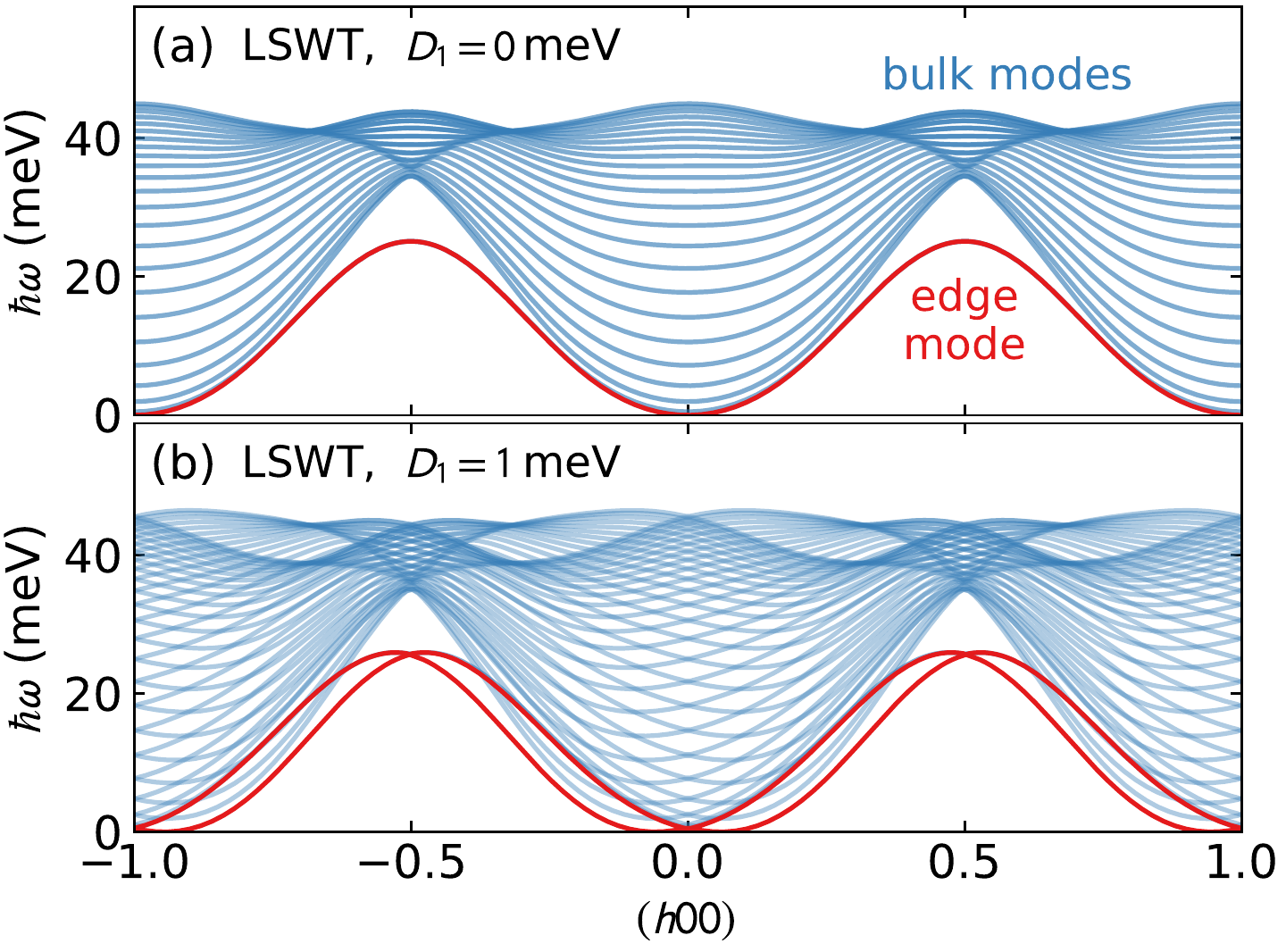}
	\caption{Large box linear spin wave theory (LSWT) simulations of CrSBr with lattice 12 unit cells along $b$. The surface magnon modes are plotted in red, while the bulk magnon modes are plotted in blue. When $D_1$ becomes nonzero, the surface modes split and have linear crossings at integer and half-integer $h$, signaling potential chiral edge modes with opposite group velocities on opposing edges.}
	\label{fig:SpinWavesSlab}
\end{figure}

To examine the effect of large DM exchange on the surface magnon modes of CrSBr, we performed large box spin wave simulations using \textit{SpinW} \cite{SpinW}. We generated a lattice 12 unit-cells in extent along the $b$ axis with periodic boundary conditions along $a$ and $c$. We then performed LSWT calculations with and without periodic boundary conditions along $b$ using the Hamiltonian in Table \ref{tab:FittedJ}, with and without $D_1$ (DM on the first neighbor) exchange. The results are plotted in Fig. \ref{fig:SpinWavesSlab}.
The surface modes are clearly visible as the modes at lower energies than the bulk dispersions, and which disappear when periodic boundary conditions are applied.

Without DM interaction, the surface modes have a sinusoidal character, with the same dispersion for both surfaces. However, with a nonzero DM interaction, the modes split and shift left and right in reciprocal space, leading to crossing points at $h=0$ and $h= \pm 1/2$ where the surface magnon modes have opposite group velocities. This signals a potentially chiral surface mode which can be induced in CrSBr. If a magnon mode is excited in the frequency and momentum window of a crossing point, its direction will be constrained by the dispersion to travel along a particular edge direction. The chiral edge modes may be visible in a thermal hall experiment. Inducing these chiral edge modes via proximity effects is a real possibility: CrSBr heterostructures are already being fabricated \cite{Ghiasi2021} and furthermore layered WTe$_2$/Fe$_3$GeTe$_2$ were able to achieve 1.0~mJ/m$^2$ proximity induced DM exchange \cite{wu2020neel}, which would be 1.9~meV per Cr ion in CrSBr---larger even than our DM simulations in Fig. \ref{fig:SpinWavesSlab}. %Thus inducing chiral edge modes via proximity effects is a very real possibility.

As an aside, these simulations show that the DM interaction would also shift the mode energy minima from $Q=0$ to an incommensurate value along $a$. This indicates that $D_1$ would produce an incommensurate spiral spin modulation along $a$ (the in-plane direction perpendicular to the ordered moment).

\section{Conclusion}

In conclusion, we have measured the magnetic diffraction of CrSBr and confirmed the 2D $XY$ A-type antiferromagnetism. We also measured inelastic spin wave spectra of CrSBr and fitted the observed magnon modes to a linear spin wave model. We find a minimal magnetic exchange model with seven in-plane exchanges accurately reproduces the experimental spectra, with both single-ion and exchange anisotropy being too small to resolve. We also find no visible dispersion in the out of plane direction, confirming the highly 2D nature of CrSBr. We anticipate this experimentally derived Hamiltonian to be useful for calculating the behavior of this material in heterostructures and spintronic devices.

We then use this calculated spin wave model to predict the presence of a chiral edge mode if the nearest neighbor DM exchange interaction could be enhanced by proximity effects. These results suggest potential topological edge modes in CrSBr heterostructures is a future direction worth exploring.

%%%%%%%%%%%%%%%%%%

\section{Experimental section}\label{sec:Experimental}

\subsection{Sample synthesis}

\paragraph{Reagents:}
 The following reagents were used as received unless otherwise stated: chromium powder (99.94\%, -200 mesh, Alfa Aesar), sulfur pieces (99.9995\%, Alfa Aesar), bromine (99.99\%, Aldrich), and chromium dichloride, (anhydrous, 99.9\%, Strem Chemicals)

\paragraph{Synthesis of CrBr$_3$:}
 High quality CrBr$_3$ was synthesized from the elements (Cr, 1.78~g, 34.2~mmol and Br$_2$, 8.41~g, 52.6~mmol) with one end of the tube maintained at $1000^{\circ}$C and the other side at ~50°C with a water bath. Details of the reaction can be found in \cite{bae2022exciton}. \textit{Caution: One end of the tube must be maintained below $120^{\circ}$C to prevent the tube from exploding from bromine overpressure.}

\paragraph{Synthesis of CrSBr:}
 A modified procedure from ref \cite{bae2022exciton} was used to synthesize large single crystals of CrSBr. Chromium (0.174~g, 3.35~mmol), sulfur (0.196~g, 6.11~mmol), and chromium tribromide (0.803~g, 2.75  mmol) were loaded into a 12.7 mm O.D., 10.5 mm I.D. fused silica tube. The tube was evacuated to a pressure of $\sim 30$~mtorr and flame sealed to a length of 20~cm. The tube was placed into a computer-controlled, two-zone, tube furnace. The source side was heated to $850^{\circ}$C in 24 hours, allowed to soak for 24 hours, heated to $950^{\circ}$C in 12 hours, allowed to soak for 48 hours, and then cooled to ambient temperature in 6 hours. The sink side was heated to $950^{\circ}$C in 24 hours, allowed to soak for 24 hours, heated to $850^{\circ}$C in 12 hours, allowed to soak for 48 hours, and then cooled to ambient temperature in 6 hours. The crystals were cleaned by soaking in a 1~mg/mL of CrCl$_2$ aqueous solution for 1 hour at ambient temperature. After soaking, the solution was decanted and the  crystals were thoroughly rinsed with DI water and acetone. Residual sulfur residue was removed by washing with warm toluene. 

\subsection{Neutron experiments}

We measured the neutron diffraction of CrSBr with the TOPAZ diffractometer \cite{coates2018suite} at Oak Ridge National Laboratory's SNS. TOPAZ uses the neutron wavelength-resolved Laue technique for data collection to measure a 3D volume from a stationary single-crystal sample. %, using a 1.4~cm long, 43~mg crystal oriented with the $a$ axis vertical. We measured the diffraction at 200~K, 80~K, and 5~K rotating the sample $180^{\circ}$; 
Diffraction study was made on a plate-shaped single crystal with dimensions 5 x 2.5 x 0.8 mm, orientated with the $a$ axis vertical. Sample temperature was controlled by a Cryomech P415 pulse tube cryocooler. Data were collected using crystal orientations optimized with the CrystalPlan software in the range $-161^{\circ}$ to $180^{\circ}$ \cite{zikovsky2011crystalplan} at 200~K, 80~K and 5~K.
We also measured an order parameter curve heating from 5~K to 200~K at a fixed rotation angle. As explained in detail in the Supplemental Information, we use the  BasIreps \cite{Fullprof} and JANA software packages \cite{Petricek_2014} to perform a refinement to the magnetic Bragg intensities and find a static ordered moment of 3.56(2)~$\mu_B$ at $T=5$~K. This is consistent with the theoretical static moment of a $S=3/2$ Cr$^{3+}$ ion: $g(3/2) = 3\>\mu_B$ plus a small orbital contribution. 

 We measured the inelastic neutron spectrum of CrSBr using the SEQUOIA spectrometer \cite{Granroth2006,Granroth2010} at Oak Ridge National Laboratory's SNS \cite{mason2006spallation}. The sample consisted of 13 coaligned crystals with a total mass of 300 mg, aligned with the $c$ axis vertical and glued to an aluminum plate using CYTOP glue \cite{rule2018glue} (a picture is shown in the Supplemental information). The sample was mounted in a closed cycle refrigerator and cooled to a base temperature of 5~K. We measured the scattering with an incident energies $E_i = 70$~meV and  $E_i = 20$~meV. %, rotating the sample 180$^{\circ}$ in 1$^{\circ}$ steps. In this way, we map out a full four-dimensional volume of reciprocal space. %, resolving $h$, $k$, $\ell$, and energy transfer $\hbar \omega$ for a four-dimensional data set.
 
 For the SEQUOIA neutron measurements, we set the $T0$ chopper at 60~Hz, and used high flux Fermi 1 chopper at 240~Hz,  for $E_i = 70$~meV, and the neutron absorbing slits in front of the sample were set to provide the beam size 44 mm wide and 6 mm tall. 
 We also measured the spectra with %$E_i = 200$~meV neutrons (high flux mode, $T0$ chopper at 120~Hz, Fermi 1 chopper at 300~Hz, Fermi 2 chopper at 120~Hz), and  
 $E_i = 20$~meV neutrons using high resolution Fermi 2 chopper at 240 Hz, $T0$ chopper at 60~Hz.
 For the  $E_i = 70$~meV % and  $E_i = 200$~meV 
 data we rotated the sample a full 180$^{\circ}$ in 1 degree steps, but for the  $E_i = 20$~meV data we rotated only 35$^{\circ}$ to capture the bottom of the dispersion around $(1,1,0)$.

%%%%%%%%%%%%%%%%%%

 \subsection*{Acknowledgments}
 A.S. acknowledges helpful discussions with Peter Holdsworth and Paul McClarty.
 This research used resources at the Spallation Neutron Source, a DOE Office of Science User Facility operated by the Oak Ridge National Laboratory. The PPMS used to perform vibrating sample magnetometry measurements was purchased with financial support from the NSF through a supplement to award DMR-1751949. 
Synthesis was supported by the National Science Foundation (NSF) Materials Research Science and Engineering Centers (MRSEC) program (DMR-2011738) and as part of Programmable Quantum Materials, an Energy Frontier Research Center funded by the U.S. Department of Energy (DOE), Office of Science, Basic Energy Sciences (BES), under award DE-SC0019443.

 \subsection*{Conflict of Interest}
 
 The authors declare no conflict of interest.

\subsection*{Data Availability}

Spin wave data is available for download at doi.org/10.13139/ORNLNCCS/1869252.

%%%%%%%%%%%%%%%%%%


\pagebreak

\quad 

\newpage

\section*{Supplemental Information for Spin waves and magnetic exchange Hamiltonian in CrSBr}

\renewcommand*{\citenumfont}[1]{S#1}
\renewcommand*{\bibnumfmt}[1]{[S#1]}
\renewcommand{\thefigure}{S\arabic{figure}}
\renewcommand{\thetable}{S\arabic{table}}
\renewcommand{\theequation}{S.\arabic{equation}}
\renewcommand{\thepage}{S\arabic{page}}  
\setcounter{figure}{0}
\setcounter{page}{1}
\setcounter{equation}{0}
\setcounter{section}{0}

\section{Diffraction and refinement}

\begin{figure*}
	\centering
	\includegraphics[width=0.6\linewidth]{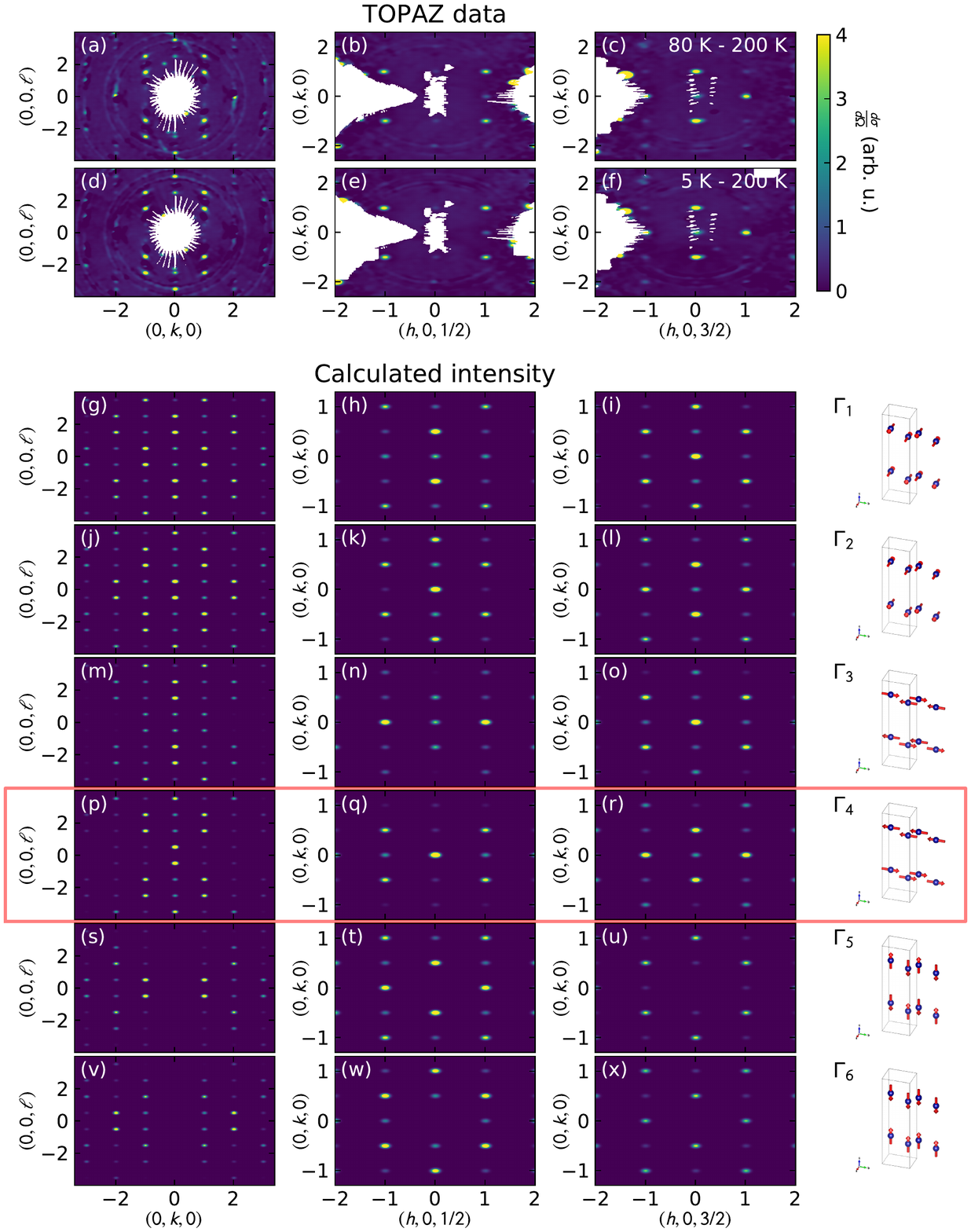}
	\caption{CrSBr single crystal diffraction and refinement. The top two rows show the experimental diffraction data, in the $(0,k,\ell)$ (left column), $(h,k,1/2)$  (middle column), $(h,k,3/2)$  (right column) scattering planes. The top row (a)-(c) shows the 80~K scattering with 200~K subtracted as background, and the second row (d)-(f) shows the same for 5~K scattering. Note that no new Bragg peaks appear at the lower temperatures. 
		The bottom section (g)-(x) shows the calculated scattering intensity for the six possible irreducible representations for CrSBr, one for each row, with the magnetic structures shown on the right. The only structure that matches the observed scattering pattern is $\Gamma_4$.}
	\label{fig:crsRefinement}
\end{figure*}

To analyze and fit the neutron diffraction peaks to a magnetic structure model, we decomposed the CrSBr space group into irreducible representations using the BasIreps package of the FullProf software suite \cite{Fullprof}. The CrSBr $Pmmn$ space group decomposes into six irreducible representations: $\Gamma_1$ -  $\Gamma_6$, shown on the right hand side of Fig. \ref{fig:crsRefinement}. Each of these six irreducible representations gives a unique signal in the magnetic diffraction pattern. Susceptibility and neutron order parameter curves indicate a second order (continuous) phase transition, which means the transition involves a single irreducible representation.

These calculated patterns are compared with the measured CrSBr scattering at 80~K and 5~K in Fig \ref{fig:crsRefinement}. Although the Bragg peaks are asymmetric and broadened due to sample imperfections, the magnetic order is nonetheless clear: the magnetic diffraction matches $\Gamma_4$, in-plane ferromagnetic order along $a$, alternating between layers to form an antiferromagnetic layered structure. 

Because no discontinuity is visible in the $(0,1,\frac{3}{2})$ order parameter curve (see main text), we believe there is no additional low-temperature transition in the magnetic order at 30~K. Nevertheless, we can use the 5~K data to constrain the possible magnetic order at lower temperatures, allowing for an additional irreducible representation which would modify the magnetic structure. Fitting combinations of irreducible representations to the observed intensities, we find that the largest possible spin canting angle at 5 K is $14^{\circ}$ to within one standard deviation uncertainty, coming from the addition of $\Gamma_1$. However, the largest possible secondary irreducible representation weight is from the addition of $\Gamma_3$ which modulates the size of the ordered moments, up to $\pm 45$\%. Thus even were a low temperature transition to exist, its effect on the overall spin structure would be quite mild. However, both of these possibilities would involve a discontinuity in the $(0,1,\frac{3}{2})$ Bragg peak, which we do not observe.

\section{Kosterlitz-Thouless physics}

\begin{figure}
	\centering
	\includegraphics[width=0.48\textwidth]{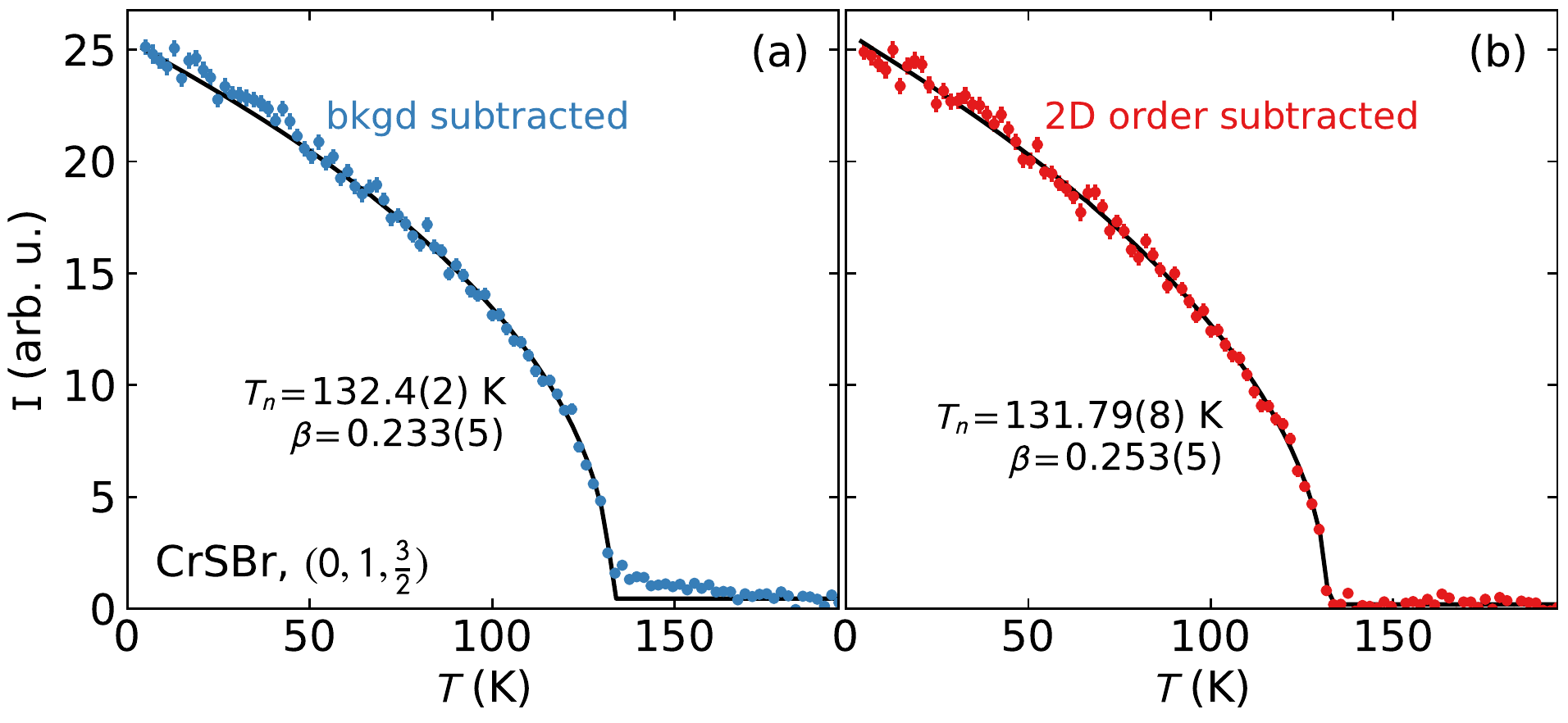}
	\caption{CrSBr order magnetic parameter measured on the $(0,1,3/2)$ Bragg peak (main text Fig. 2), with the background (a) and with 2D order parameter (b) subtracted. The fitted critical exponent with the background subtracted data agrees with the main text $\beta=0.231(6)$ to within uncertainty, but the fitted exponent with 2D order subtracted is noticeably higher. Intuitively, this makes sense: upon subtracting the 2D correlations, the result is more three-dimensional.}
	\label{fig:orderparameter03}
\end{figure}

In the main text we show that the magnetic order parameter follows the 2D $XY$ critical exponent associated with Kosterlitz-Thouless (K-T) physics \cite{Bramwell_1993}, indicating the possible existence of magnetic vortices in the paramagnetic phase. As shown in Fig. \ref{fig:orderparameter03}, this result is robust against background subtracted fits.

\begin{figure}
	\centering
	\includegraphics[width=0.38\textwidth]{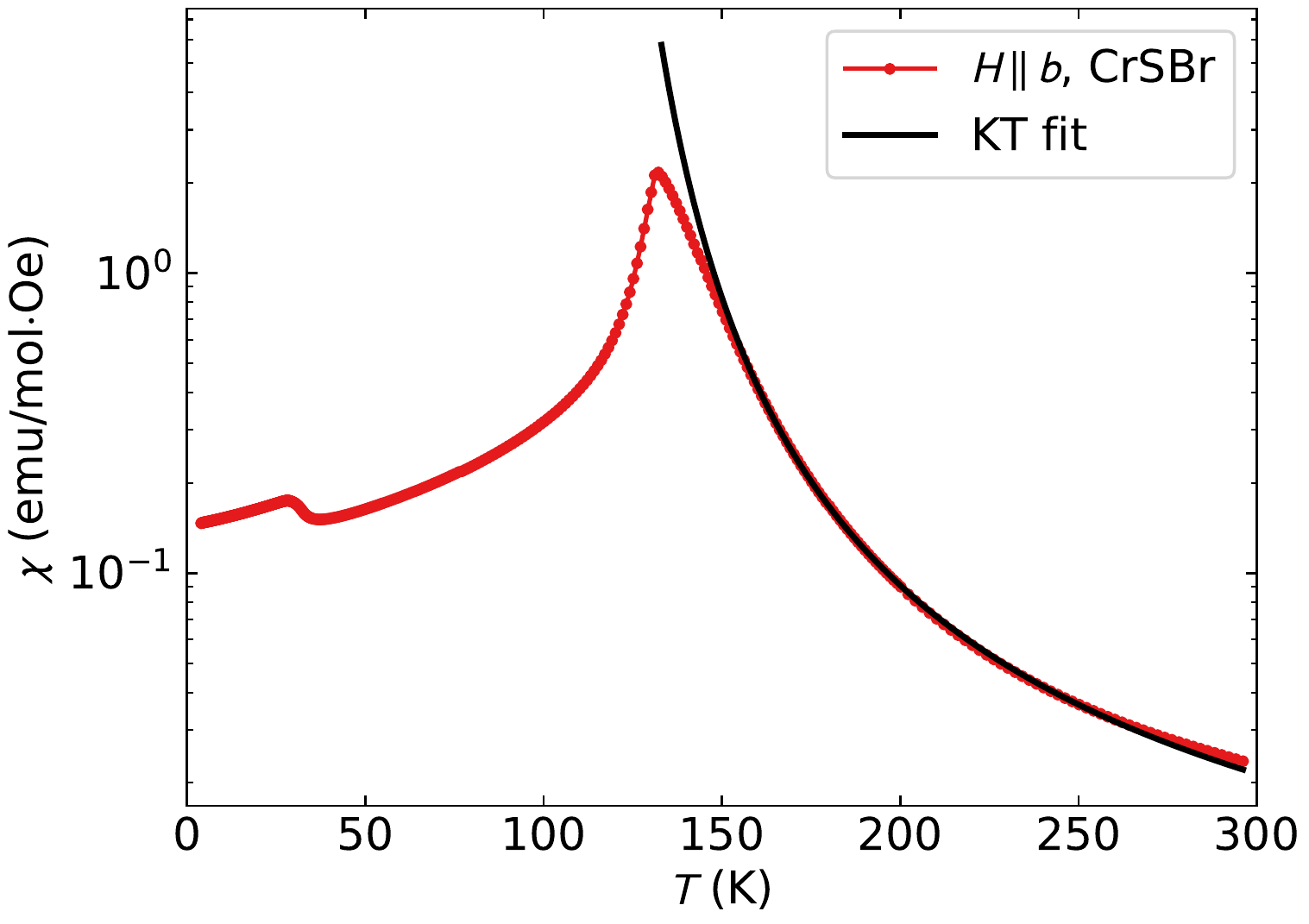}
	\caption{CrSBr magnetic susceptibility along $b$, from Ref. \cite{telford2020layered}, fitted to Eq. \ref{eq:KTsus} above the magnetic ordering transition. Between $\sim 150$~K and 300~K, the susceptibility follows the K-T functional form.}
	\label{fig:KTfit}
\end{figure}

One of the original predictions of Kosterlitz \cite{Kosterlitz_1974,Cornelius_1986} was that the high temperature susceptibility of a K-T system follows a universal curve 
\begin{equation}
\chi(T) = A \exp(B \> t^{-1/2}) \label{eq:KTsus}
\end{equation}
where $A$ and $B$ are fitted constants and $t=(T-T_{KT})/T_{KT}$. Here $T_{KT}$ is the vortex binding transition, below which no free vortices can exist in the lattice. Fitting the high temperature CrSBr susceptibility to this formula, we obtain very good agreement down to $\sim 150$~K as shown in Fig. \ref{fig:KTfit}, with a fitted  $T_{KT}=106.3(5)$~K in the $b$ and $c$ directions and $T_{KT}=93.7(3)$~K in the $a$ direction. This confirms that CrSBr behaves as a 2D K-T system, and potentially that around 100~K magnetic vortices and antivortices would bind together and annihilate. Intriguingly, the fitted $T_{KT}$ transition is nearly the same as the $T_S$ temperature identified by muon spin relaxation \cite{lopez2022dynamic}. Whether this is coincidence or indicates a deep connection is a question left for future study.

\section{SEQUOIA Experiment details}

\begin{figure}
	\centering\includegraphics[width=0.25\textwidth]{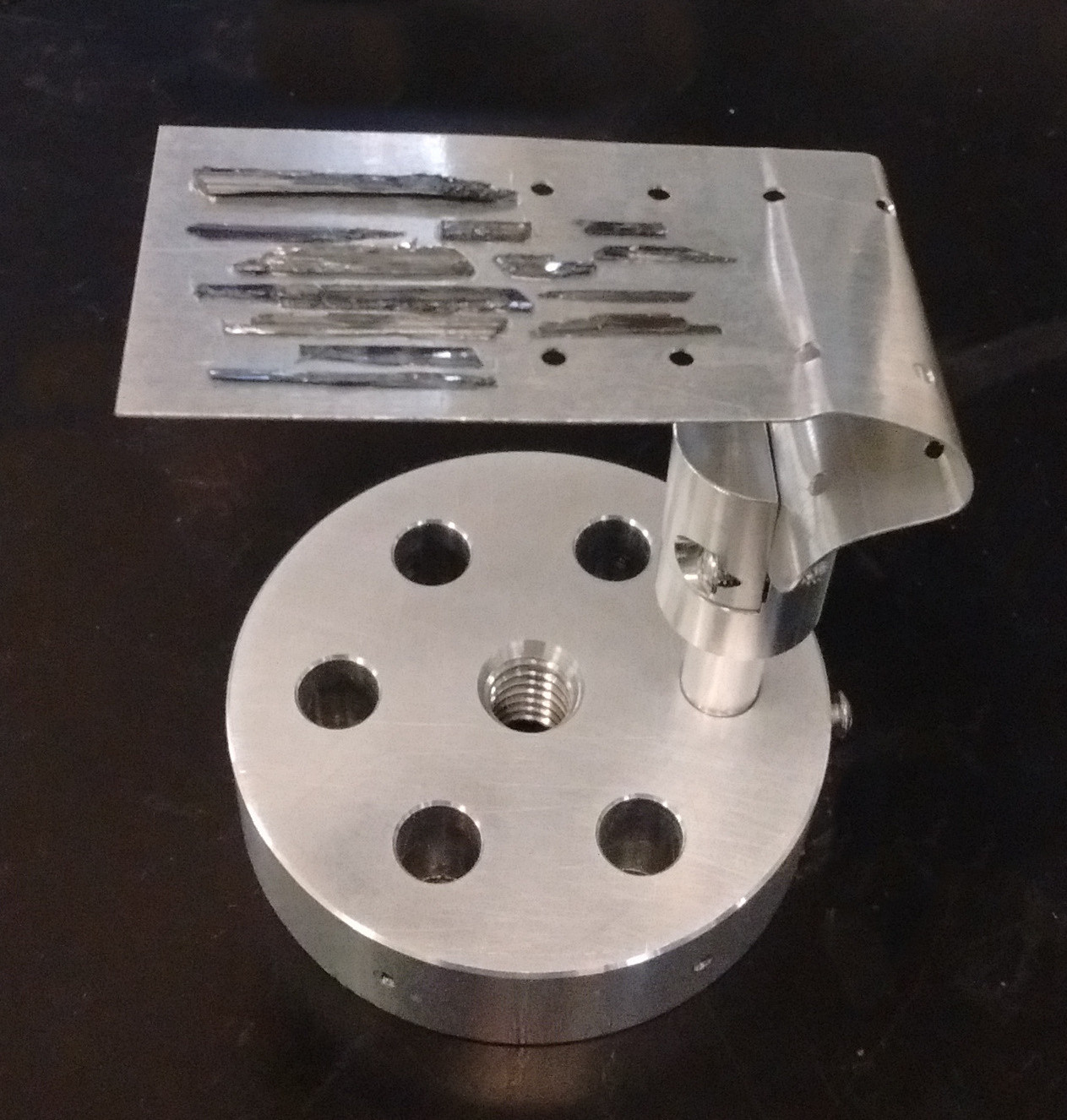}
	\caption{CrSBr sample used for the neutron experiment, consisting of 13 coaligned crystals glued to an aluminum plate for a total mass of 300 mg. The $c$ axis is vertical, and the longest crystal edge is along the $a$ axis.}
	\label{fig:sample}
\end{figure}

The sample for this measurement is shown in Fig. \ref{fig:sample}. It consists of 13 crystals coaligned with the $c$ axis vertical, glued to an aluminum plate with CYTOP glue \cite{rule2018glue}. 

\begin{figure}
	\centering\includegraphics[width=0.48\textwidth]{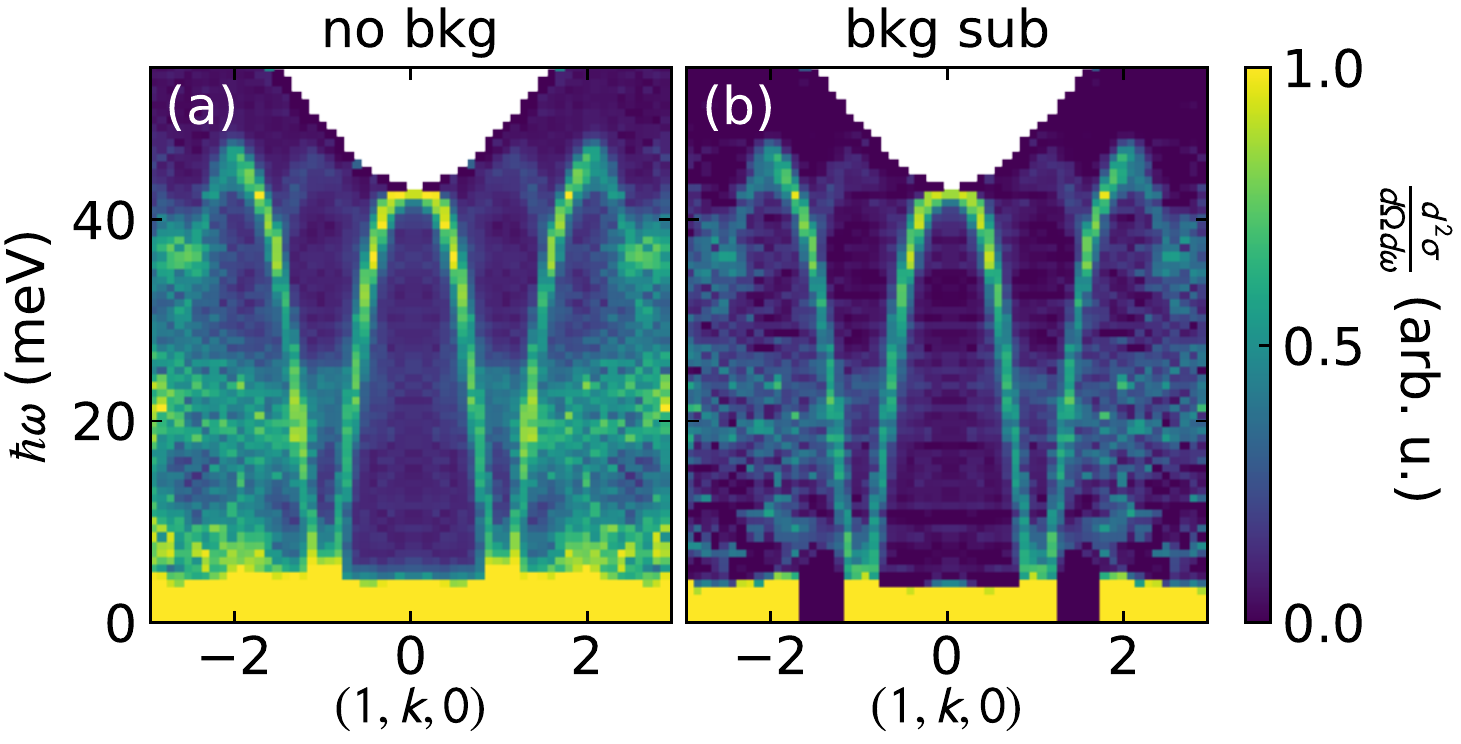}
	\caption{CrSBr scattering with and without the background subtracted. The background significantly reduces the intensity of the diffuse features whilst leaving the magnon intensity unchanged.}
	\label{fig:background}
\end{figure}

To estimate background for this experiment, we moved the absorbing slits up by 9 mm so that the neutron beam illuminates the middle of the aluminum sample mount and not the CrSBr crystals. This was subtracted from the data as shown in Fig. \ref{fig:background}. The diffuse background is not eliminated, but it is reduced to the point where it is clear what the magnon modes are.

\begin{figure}
	\centering\includegraphics[width=0.48\textwidth]{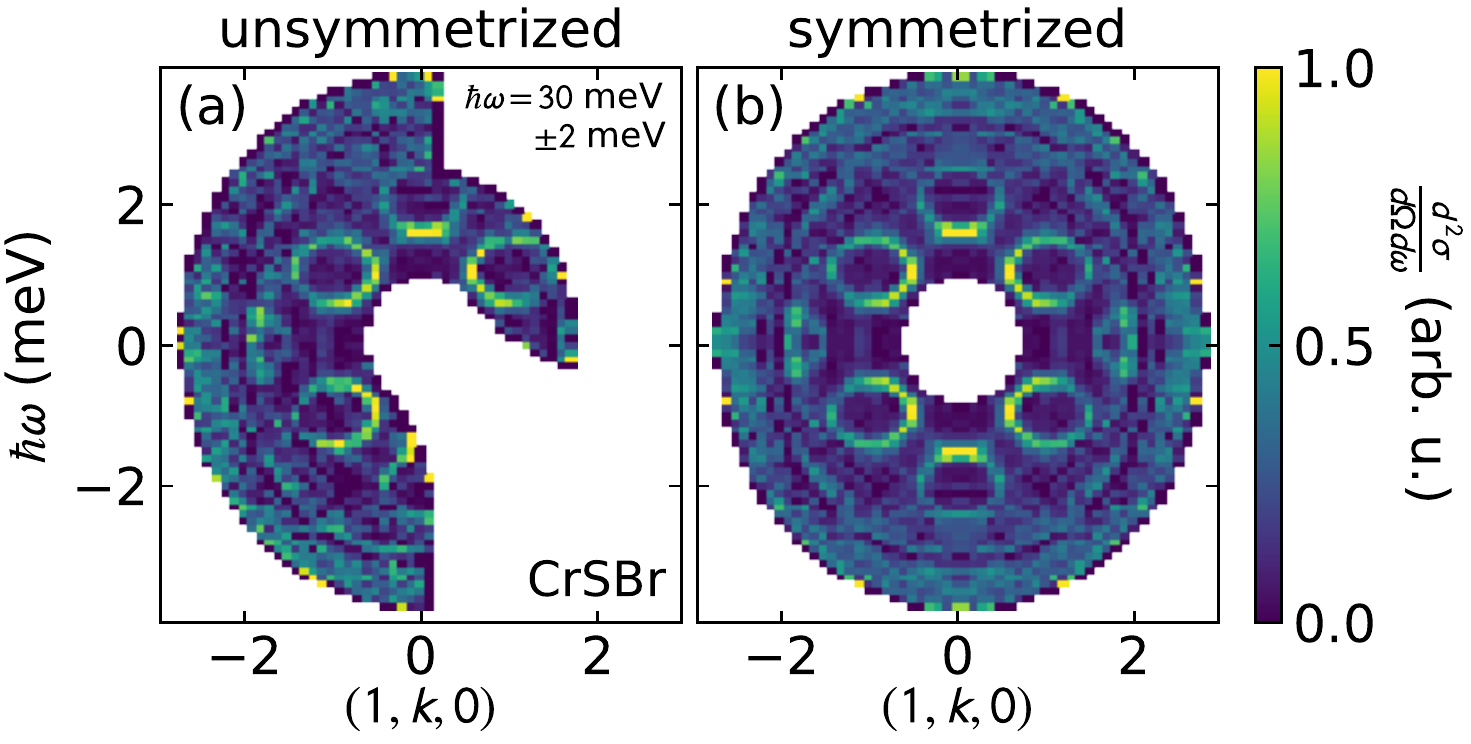}
	\caption{Symmetrized and unsymmetrized CrSBr data, showing a constant energy slice at $\hbar \omega = (30 \pm 2)$~meV.}
	\label{fig:symmetrization}
\end{figure}

To increase the statistical clarity of the CrSBr scattering data, we symmetrized the data by folding it over high-symmetry directions as shown in Fig. \ref{fig:symmetrization}. The magnon modes are clear even without symmetrizing, but applying the symmetry operations increases the effective signal intensity.

\section{Spin wave fits}\label{app:fits}

As noted in the main text, the spin wave Hamiltonian was fitted to the energies of magnon modes extracted from the scattering data. This optimization was repeated several times using different starting parameters, and always converged to the same solution.

\begin{figure*}
	\centering
	\includegraphics[width=0.78\textwidth]{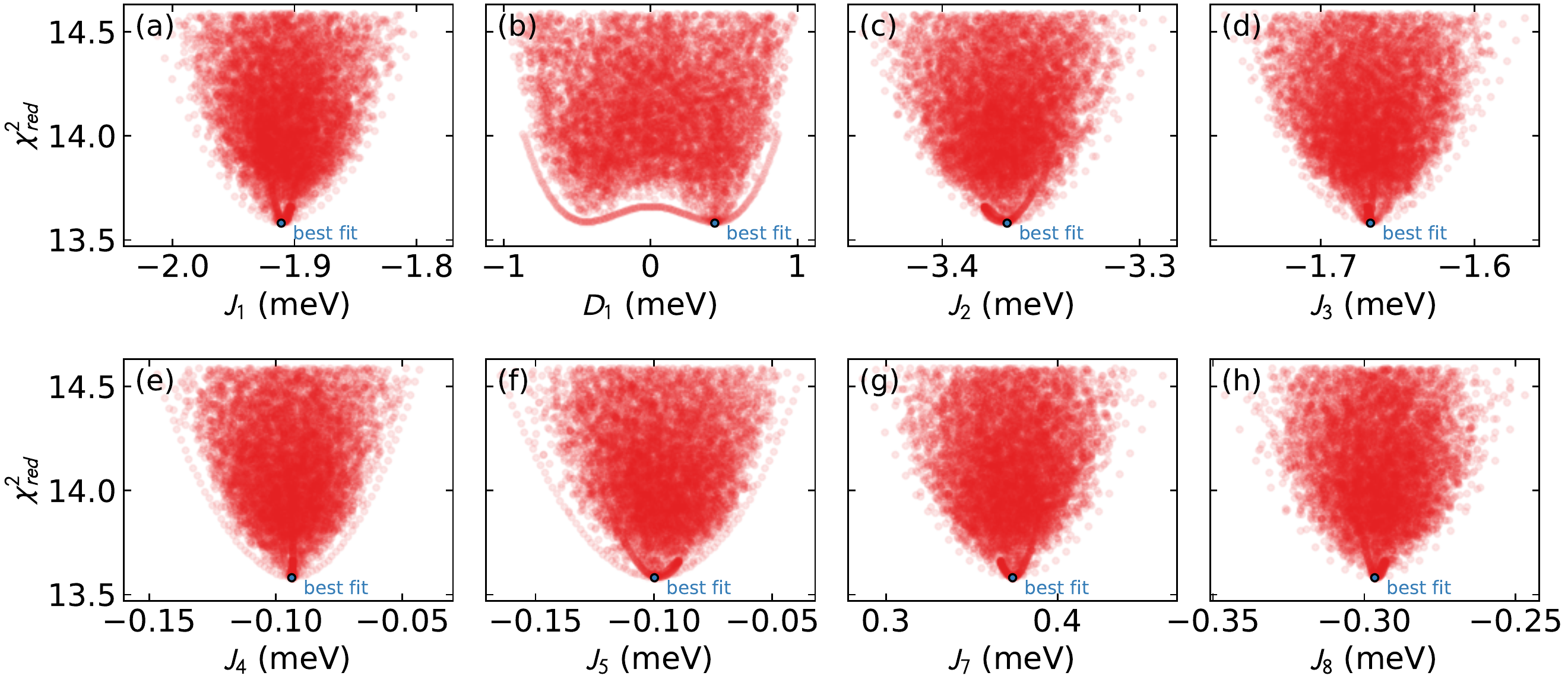}
	\caption{Possible Hamiltonian solutions within $\Delta \chi^2_{red}=1$ of the global optimum fit (see text). Each panel shows the range of such solutions, which we take as an estimate of uncertainty. The small blue circle represents the best fit values. Note that the DM interaction in panel (b) is symmetric about $D_1=0$.}
	\label{flo:jrangeplot}
\end{figure*}

Uncertainty was estimated by calculating the $\Delta \chi^2_{red}=1$ contour about the best fit Hamiltonian. The extremities of this contour along each parameter is then a measure of the one standard deviation uncertainty \cite{NumericalRecipes}.
We estimate the $\Delta \chi^2_{red}=1$ using the same method as employed in Ref. \cite{scheie2022spin}, by generating a family of solutions within $\Delta \chi^2_{red}=1$ of the best fit Hamiltonian. We first systematically vary a single parameter, slowly increasing or decreasing its value while re-fitting all other values, keeping solutions within $\Delta \chi^2_{red} \leq 1$. Then, using principal component analysis to define the vectors along parameter space to search, we perform a random Monte Carlo search to generate additional solutions within $\Delta \chi^2_{red}=1$. The ranges of valid solutions are plotted in Fig. \ref{flo:jrangeplot}, which are used to define the statistical uncertainty in main text Table I.

\begin{figure*}
	\centering
	\includegraphics[width=0.78\textwidth]{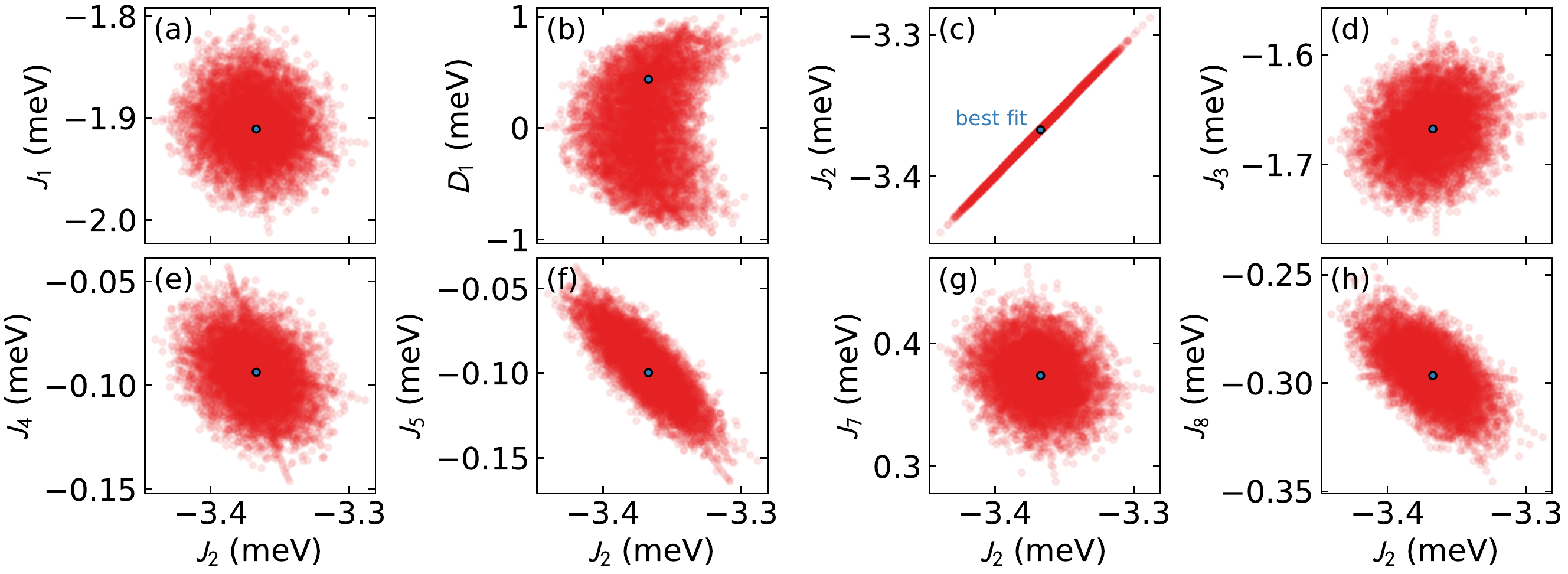}
	\caption{Hamiltonian solutions within $\Delta \chi^2_{red}=1$ plotted against $J_2$, as an example of correlations between fitted parameters. A circular distribution shows no correlation, but a tilted ellipsoidal distribution shows nonzero correlation. The small blue circle represents the best fit values.}
	\label{flo:correlationjplot}
\end{figure*}

\begin{figure}
	\centering\includegraphics[width=0.38\textwidth]{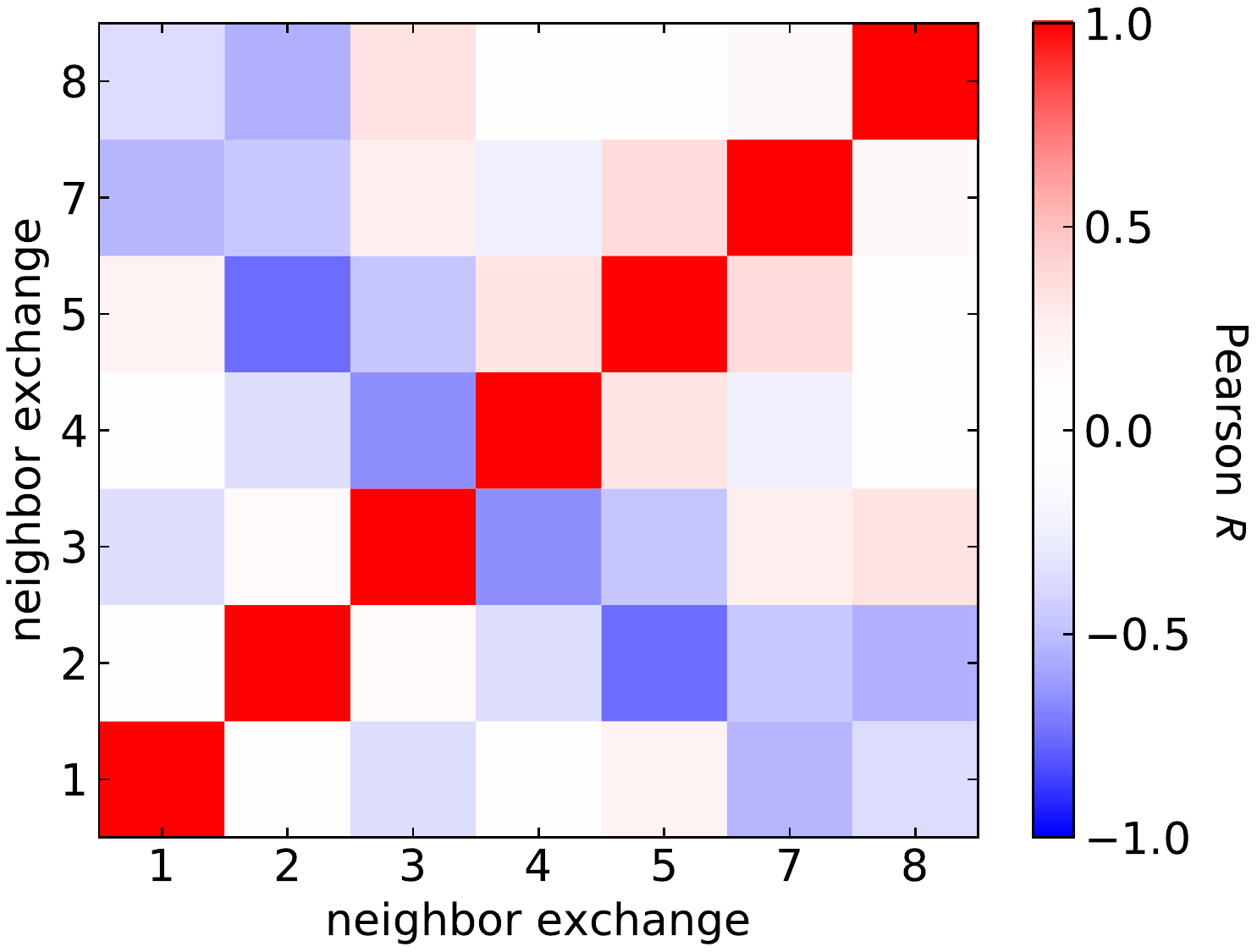}
	\caption{Correlation matrix for nonzero $J_n$ from the CrSBr spin wave mode fits, generated from the data in Fig. \ref{flo:correlationjplot} via a Pearson $R$ calculation. Red indicates positive correlation, blue indicates negative correlation.}
	\label{fig:CorrelationMatrix}
\end{figure}

This method of calculating uncertainty also yields the correlations between the various fitted parameters. Fig. \ref{flo:correlationjplot} shows the correlations between different parameters on $J_2$ in the best fit family of solutions. This is then used to calculate a correlation matrix, as shown in Fig. \ref{fig:CorrelationMatrix}.

The case of the DM exchange $D_1$ is somewhat unique as it effects only some of the dispersion $Q$ points. (It leaves integer and half-integer $h$ unchanged for instance.) So for estimating the uncertainty of $D_1$, we modified $\chi^2_{red}$ to be $\chi^2$ per data point for which DM has a noticeable effect. This reduced the uncertainty from $\pm 0.5$~meV to $\pm 0.4$~meV. But in either case, the uncertainty in $D_1$ overlaps with zero.

\begin{figure}
	\centering\includegraphics[width=0.4\textwidth]{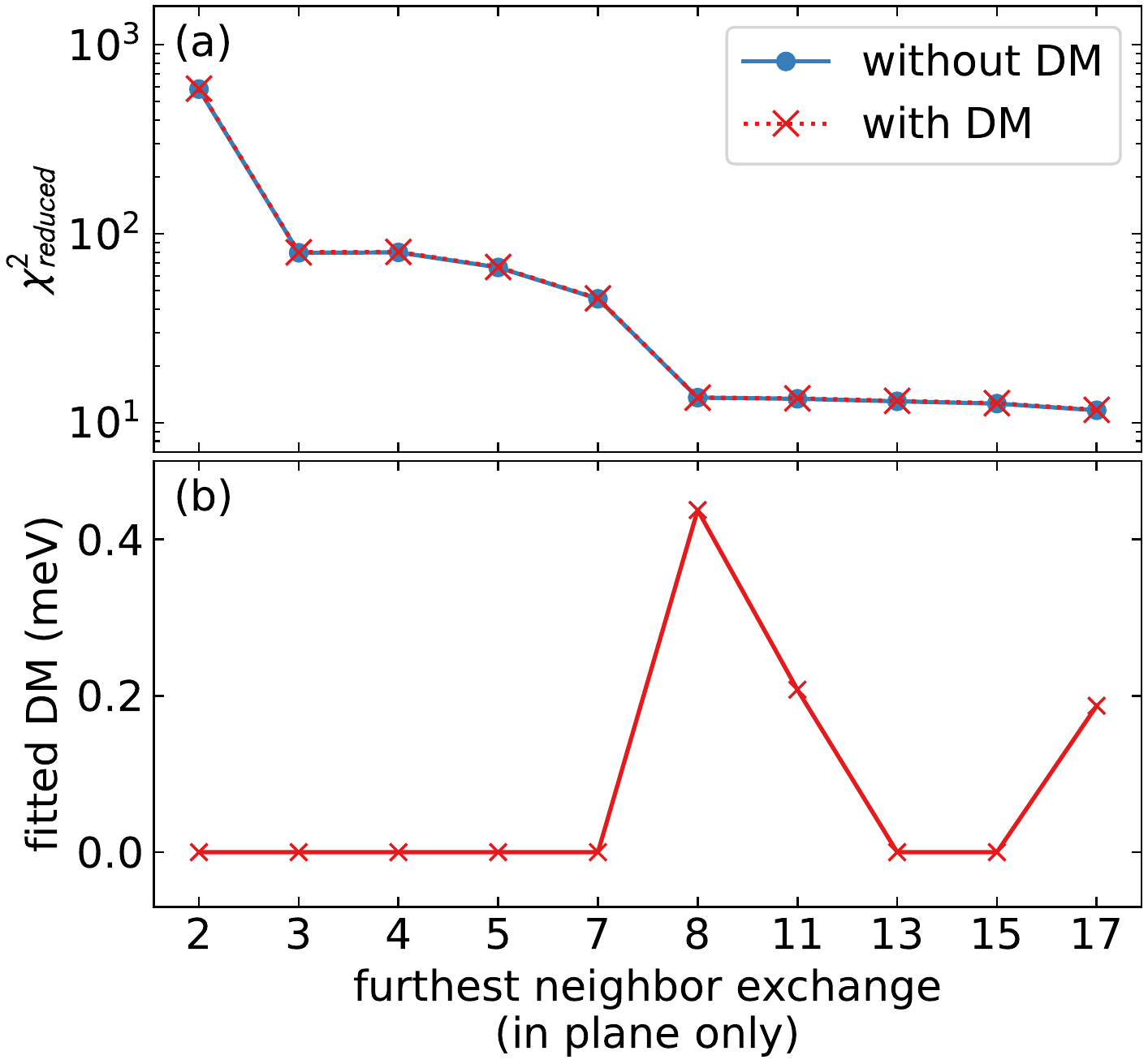}
	\caption{(a) dependence of the best fit $\chi_{\rm red}^2$ on the number of neighbors $n$, both with and without a DM term included. The overall $\chi^2$ is barely different when $D_1$ is included, and in fact is slightly higher for all $n$. (b) Best fit $D_1$ value plotted against $n$, showing the fitted value is unstable with $n$. Thus the DM cannot be accurately determined by this data, but is small enough that it makes no difference to fit.}
	\label{fig:Chisq_vs_n_DM}
\end{figure}

Furthermore, although with $n=8$ neighbors included in the fit (what is plotted in Fig. \ref{flo:jrangeplot}) gives a best fit value of $D_1 \approx 0.4$, this value is unstable with $n$. As shown in Fig. \ref{fig:Chisq_vs_n_DM}, the best fit $D_1$ value varies unpredictably as the number of fitted $J_n$ increases---and in most cases refines to zero. 
Additionally, the overall $\chi^2_{red}$ is never improved by adding a $D_1$ term to the Hamiltonian, as shown in Fig. \ref{fig:Chisq_vs_n_DM}(a). 
Because of this, we cannot say that $D_1$ is nonzero in CrSBr, but based its statistical uncertainty it could be as large as 0.8~meV.

\section{Inter-plane exchange}\label{app:interplane}

\begin{figure*}
	\centering
	\includegraphics[width=0.7\textwidth]{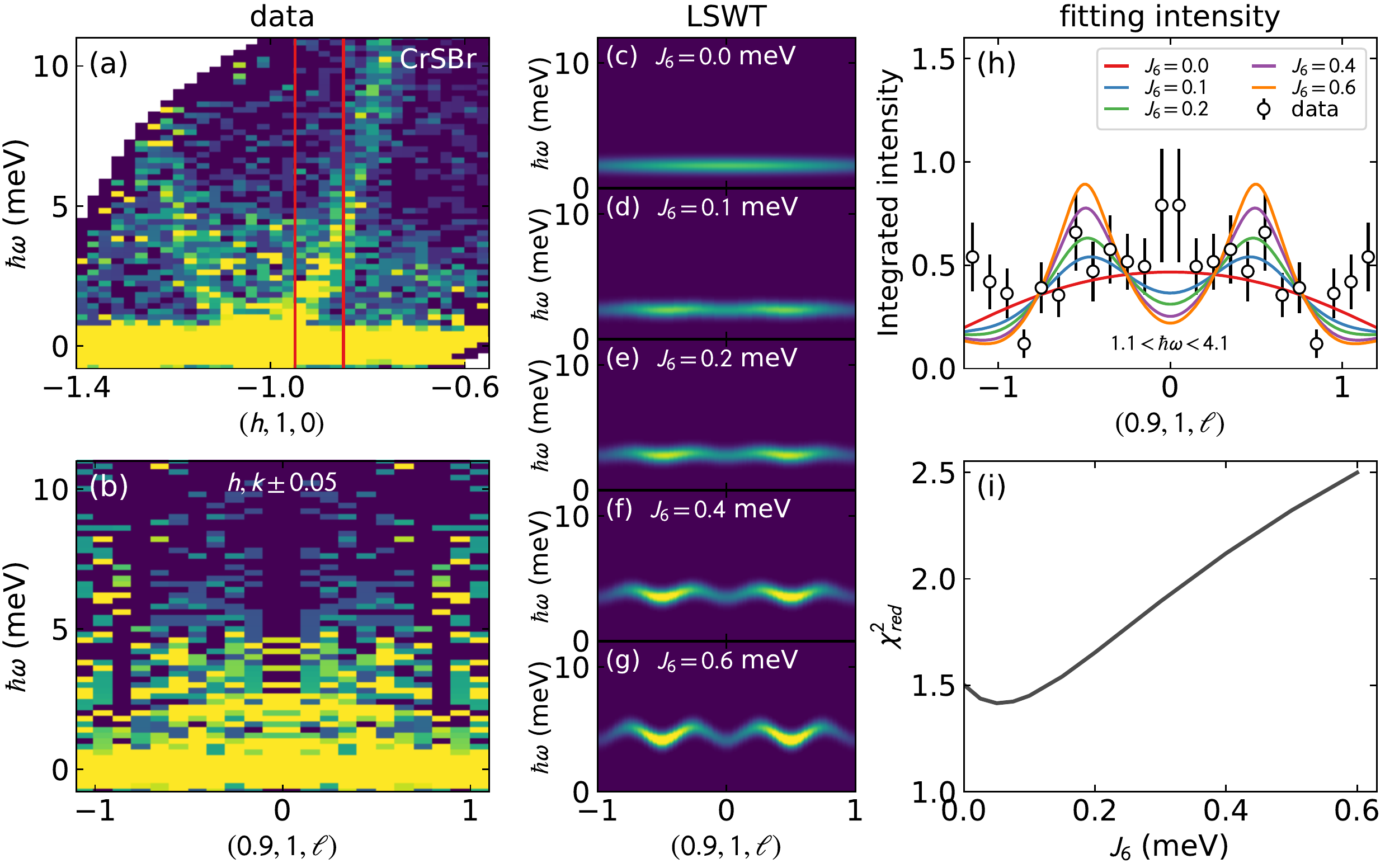}
	\caption{Fitted dispersion in the out of plane ($\ell$) direction. (a) shows $E_i=20$~meV data, with the red lines indicating the integrated region in panel (b). Panels (c) through (g) show linear spin wave simulations of the CrSBr dispersion with varying values of the inter-plane exchange $J_6$. As the interplane exchange grows larger, the intensity concentrates around $\ell = \pm 1/2$. Panel (h) shows the integrated intensity from panel (b) in the window between 1.1~meV and 4.1 meV compared to the linear spin wave simulations. Panel (i) shows the reduced $\chi^2$ as a function of $J_6$, giving a best fit of $J_6=0.05$~meV, with an uncertainty up to 0.5~meV.}
	\label{flo:interplane}
\end{figure*}

As noted in the main text, no inter-plane magnetic dispersion is detectable to within the resolution of the neutron experiment. Nevertheless, we can attempt to put an upper bound on the $J_6$ inter-plane exchange by examining the lowest energy inelastic scattering, as shown in Fig. \ref{flo:interplane}. 

As is shown by the LSWT calculations in Fig. \ref{flo:interplane}, small nonzero  $J_6$ causes the low energy magnon spectral weight to concentrate around $\ell = \pm 1/2$. Experimentally, no such intensity concentration is observed. Fitting the $\ell$ dependence of the intensity at $h=0.9$, $k=1.0$ (displaced from the $(1,1,\ell)$ so that the inelastic magnetic signal is distinguishable from the elastic background) in Fig. \ref{flo:interplane}(h), we find that the best fit  $J_6$ is 0.05~meV with an uncertainty of 0.5~meV (estimated from a one standard deviation in reduced $\chi^2$). The error bars are unfortunately too large to constrain the fit very well, but it is clear that $J_6$ is quite small in CrSBr, and the system acts very two-dimensionally.

\section{Magnetic Anisotropy}

\begin{figure}
	\centering\includegraphics[width=0.49\textwidth]{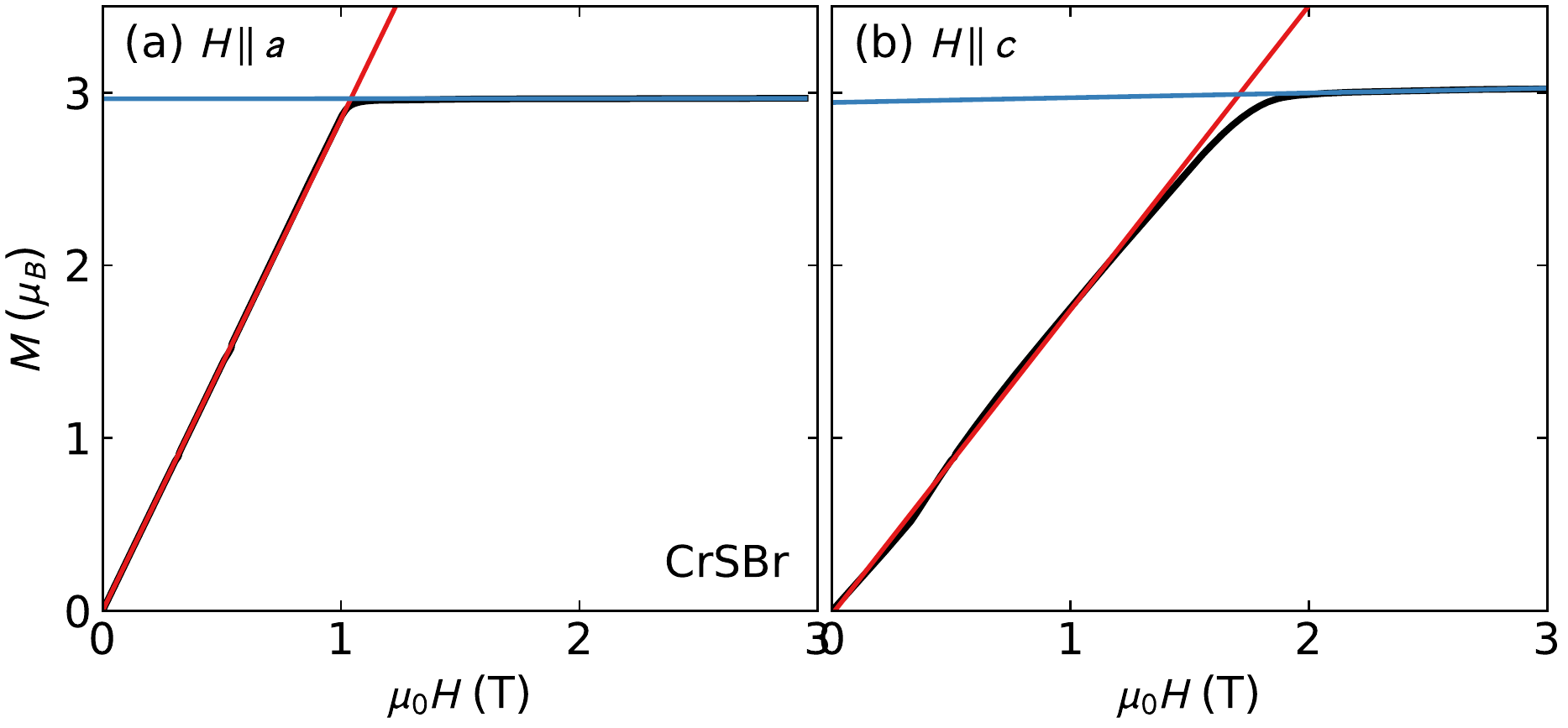}
	\caption{CrSBr 2~K magnetization as a function of field in the (a) $a$ direction and (b) $c$ direction. The black curve indicates the experimental data, the red and blue lines are linear fits.}
	\label{fig:magnetization}
\end{figure}

Although the neutron experiments are unable to resolve the low energy magnon gap, we can estimate the magnetic anisotropy using isothermal magnetization. 
For magnetocrystalline anisotropy calculations, a single crystal was oriented along the $a$ or $c$ axis and attached to a quartz paddle with GE varnish, and field-dependent magnetization measurements were collected at 2~K using the vibrating sample magnetometry module of a Quantum Design PPMS Dynacool system. The same crystal was used for both $a$- and $c$-axis measurements. The magnetocrystalline anisotropy (defined here as the energy difference between the $b$-axis-magnetized state and the $a$- or $c$-axis magnetized state) was calculated using the Stoner-Wohlfarth model \cite{stoner1948mechanism}. %Two linear fits were applied to the magnetization data: One in the low-field regime and one-in the high-field regime where the magnetization is saturated. The intersection of these two lines affords the saturation magnetization as its ordinate and the saturation field as its abscissa.
By measuring the magnetization as a function of field at 2~K, we estimate the saturation magnetic field $H_{sat}$ and magnetization $M_{sat}$ by linearly fitting the low field and high field magnetization and finding the intersection as shown in Fig. \ref{fig:magnetization}. By calculating the effective anisotropy parameter $K_{eff} = (\mu_0 H_{sat}M_{sat})/2$ for each direction as in Refs. \cite{zhdanova2011magnetocrystalline,Zhdanova2013}, we can find the difference between the anisotropy constants for the $a$ or $c$ axis and the $b$-axis. These are 144~$\mu$eV/Cr ($c$-axis compared to $b$) and 90~$\mu$eV/Cr ($a$-axis compared to $b$). These estimates are remarkably close to the measured magnon gaps from transient reflectance spectroscopy of 0.141(4)~meV and 0.102(3)~meV \cite{bae2022exciton}. Taken as estimates of the spin wave gap, these are much to small to be resolved with the reported neutron data. Thus, on short timescales (high frequencies), CrSBr can be approximated as a very isotropic 2D magnet.

We note that the values calculated here do not account for possible effects of shape anisotropy, which would be largest for the $c$-axis, given the plate-like morphology of the crystals. However, correcting the magnetization data with a demagnetization factor near unity ($N = 0.9$) alters the calculated anisotropy energy for the $c$-axis by less than 10\%. As such, possible effects of shape anisotropy do not alter any conclusions discussed herein.

%\section{Dispersion plots}
%
%
%Figures \ref{flo:SWfit_0} through \ref{flo:SWfit_13} show the measured dispersion compared to the linear spin wave theory (LSWT) simulated intensity from the best fit Hamiltonian in the main text.
%Note that the data were integrated over a finite region in $\ell$, rendering the optic modes faintly visible whereas at $\ell=0$ they would have zero intensity. To account for this, the LSWT simulations were carried out for $\ell=1/2$.
%The third panel on the right shows line plots of the calculated dispersion curves. If experimental data were extracted from the particular slice and used to constrain the fit, they are plotted overtop the calculated dispersion. In some cases, the data were too noisy to extract the modes, but one can see that the calculated magnon intensity matches the features in the data.
%
%
%\foreach \n in {0,...,13}{
%	
%	\begin{figure*}
%		%\centering
%		\includegraphics[width=\linewidth]{figures/FitPlotComparisons/CrSBr_SpinWave_Cut_\n.pdf} 
%		\caption{Experimental CrSBr scattering compared to LSWT fit.}
%		\label{flo:SWfit_\n}
%	\end{figure*}
%	
% }



\begin{thebibliography}{43}%
	\makeatletter
	\providecommand \@ifxundefined [1]{%
		\@ifx{#1\undefined}
	}%
	\providecommand \@ifnum [1]{%
		\ifnum #1\expandafter \@firstoftwo
		\else \expandafter \@secondoftwo
		\fi
	}%
	\providecommand \@ifx [1]{%
		\ifx #1\expandafter \@firstoftwo
		\else \expandafter \@secondoftwo
		\fi
	}%
	\providecommand \natexlab [1]{#1}%
	\providecommand \enquote  [1]{``#1''}%
	\providecommand \bibnamefont  [1]{#1}%
	\providecommand \bibfnamefont [1]{#1}%
	\providecommand \citenamefont [1]{#1}%
	\providecommand \href@noop [0]{\@secondoftwo}%
	\providecommand \href [0]{\begingroup \@sanitize@url \@href}%
	\providecommand \@href[1]{\@@startlink{#1}\@@href}%
	\providecommand \@@href[1]{\endgroup#1\@@endlink}%
	\providecommand \@sanitize@url [0]{\catcode `\\12\catcode `\$12\catcode
		`\&12\catcode `\#12\catcode `\^12\catcode `\_12\catcode `\%12\relax}%
	\providecommand \@@startlink[1]{}%
	\providecommand \@@endlink[0]{}%
	\providecommand \url  [0]{\begingroup\@sanitize@url \@url }%
	\providecommand \@url [1]{\endgroup\@href {#1}{\urlprefix }}%
	\providecommand \urlprefix  [0]{URL }%
	\providecommand \Eprint [0]{\href }%
	\providecommand \doibase [0]{https://doi.org/}%
	\providecommand \selectlanguage [0]{\@gobble}%
	\providecommand \bibinfo  [0]{\@secondoftwo}%
	\providecommand \bibfield  [0]{\@secondoftwo}%
	\providecommand \translation [1]{[#1]}%
	\providecommand \BibitemOpen [0]{}%
	\providecommand \bibitemStop [0]{}%
	\providecommand \bibitemNoStop [0]{.\EOS\space}%
	\providecommand \EOS [0]{\spacefactor3000\relax}%
	\providecommand \BibitemShut  [1]{\csname bibitem#1\endcsname}%
	\let\auto@bib@innerbib\@empty
	%</preamble>
	\bibitem [{\citenamefont {Huang}\ \emph {et~al.}(2017)\citenamefont {Huang},
		\citenamefont {Clark}, \citenamefont {Navarro-Moratalla}, \citenamefont
		{Klein}, \citenamefont {Cheng}, \citenamefont {Seyler}, \citenamefont
		{Zhong}, \citenamefont {Schmidgall}, \citenamefont {McGuire}, \citenamefont
		{Cobden}, \citenamefont {Yao}, \citenamefont {Xiao}, \citenamefont
		{Jarillo-Herrero},\ and\ \citenamefont {Xu}}]{Huang2017}%
	\BibitemOpen
	\bibfield  {author} {\bibinfo {author} {\bibfnamefont {B.}~\bibnamefont
			{Huang}}, \bibinfo {author} {\bibfnamefont {G.}~\bibnamefont {Clark}},
		\bibinfo {author} {\bibfnamefont {E.}~\bibnamefont {Navarro-Moratalla}},
		\bibinfo {author} {\bibfnamefont {D.~R.}\ \bibnamefont {Klein}}, \bibinfo
		{author} {\bibfnamefont {R.}~\bibnamefont {Cheng}}, \bibinfo {author}
		{\bibfnamefont {K.~L.}\ \bibnamefont {Seyler}}, \bibinfo {author}
		{\bibfnamefont {D.}~\bibnamefont {Zhong}}, \bibinfo {author} {\bibfnamefont
			{E.}~\bibnamefont {Schmidgall}}, \bibinfo {author} {\bibfnamefont {M.~A.}\
			\bibnamefont {McGuire}}, \bibinfo {author} {\bibfnamefont {D.~H.}\
			\bibnamefont {Cobden}}, \bibinfo {author} {\bibfnamefont {W.}~\bibnamefont
			{Yao}}, \bibinfo {author} {\bibfnamefont {D.}~\bibnamefont {Xiao}}, \bibinfo
		{author} {\bibfnamefont {P.}~\bibnamefont {Jarillo-Herrero}},\ and\ \bibinfo
		{author} {\bibfnamefont {X.}~\bibnamefont {Xu}},\ }\bibfield  {title}
	{\bibinfo {title} {Layer-dependent ferromagnetism in a van der {Waals}
			crystal down to the monolayer limit},\ }\href
	{https://doi.org/10.1038/nature22391} {\bibfield  {journal} {\bibinfo
			{journal} {Nature}\ }\textbf {\bibinfo {volume} {546}},\ \bibinfo {pages}
		{270} (\bibinfo {year} {2017})}\BibitemShut {NoStop}%
	\bibitem [{\citenamefont {Gong}\ \emph {et~al.}(2017)\citenamefont {Gong},
		\citenamefont {Li}, \citenamefont {Li}, \citenamefont {Ji}, \citenamefont
		{Stern}, \citenamefont {Xia}, \citenamefont {Cao}, \citenamefont {Bao},
		\citenamefont {Wang}, \citenamefont {Wang} \emph
		{et~al.}}]{gong2017discovery}%
	\BibitemOpen
	\bibfield  {author} {\bibinfo {author} {\bibfnamefont {C.}~\bibnamefont
			{Gong}}, \bibinfo {author} {\bibfnamefont {L.}~\bibnamefont {Li}}, \bibinfo
		{author} {\bibfnamefont {Z.}~\bibnamefont {Li}}, \bibinfo {author}
		{\bibfnamefont {H.}~\bibnamefont {Ji}}, \bibinfo {author} {\bibfnamefont
			{A.}~\bibnamefont {Stern}}, \bibinfo {author} {\bibfnamefont
			{Y.}~\bibnamefont {Xia}}, \bibinfo {author} {\bibfnamefont {T.}~\bibnamefont
			{Cao}}, \bibinfo {author} {\bibfnamefont {W.}~\bibnamefont {Bao}}, \bibinfo
		{author} {\bibfnamefont {C.}~\bibnamefont {Wang}}, \bibinfo {author}
		{\bibfnamefont {Y.}~\bibnamefont {Wang}}, \emph {et~al.},\ }\bibfield
	{title} {\bibinfo {title} {Discovery of intrinsic ferromagnetism in
			two-dimensional van der {Waals} crystals},\ }\href
	{https://doi.org/10.1038/nature22060} {\bibfield  {journal} {\bibinfo
			{journal} {Nature}\ }\textbf {\bibinfo {volume} {546}},\ \bibinfo {pages}
		{265} (\bibinfo {year} {2017})}\BibitemShut {NoStop}%
	\bibitem [{\citenamefont {Burch}\ \emph {et~al.}(2018)\citenamefont {Burch},
		\citenamefont {Mandrus},\ and\ \citenamefont {Park}}]{Burch2018}%
	\BibitemOpen
	\bibfield  {author} {\bibinfo {author} {\bibfnamefont {K.~S.}\ \bibnamefont
			{Burch}}, \bibinfo {author} {\bibfnamefont {D.}~\bibnamefont {Mandrus}},\
		and\ \bibinfo {author} {\bibfnamefont {J.-G.}\ \bibnamefont {Park}},\
	}\bibfield  {title} {\bibinfo {title} {Magnetism in two-dimensional van der
			{W}aals materials},\ }\href {https://doi.org/10.1038/s41586-018-0631-z}
	{\bibfield  {journal} {\bibinfo  {journal} {Nature}\ }\textbf {\bibinfo
			{volume} {563}},\ \bibinfo {pages} {47} (\bibinfo {year} {2018})}\BibitemShut
	{NoStop}%
	\bibitem [{\citenamefont {Chen}\ \emph {et~al.}(2018)\citenamefont {Chen},
		\citenamefont {Chung}, \citenamefont {Gao}, \citenamefont {Chen},
		\citenamefont {Stone}, \citenamefont {Kolesnikov}, \citenamefont {Huang},\
		and\ \citenamefont {Dai}}]{Chen_2018}%
	\BibitemOpen
	\bibfield  {author} {\bibinfo {author} {\bibfnamefont {L.}~\bibnamefont
			{Chen}}, \bibinfo {author} {\bibfnamefont {J.-H.}\ \bibnamefont {Chung}},
		\bibinfo {author} {\bibfnamefont {B.}~\bibnamefont {Gao}}, \bibinfo {author}
		{\bibfnamefont {T.}~\bibnamefont {Chen}}, \bibinfo {author} {\bibfnamefont
			{M.~B.}\ \bibnamefont {Stone}}, \bibinfo {author} {\bibfnamefont {A.~I.}\
			\bibnamefont {Kolesnikov}}, \bibinfo {author} {\bibfnamefont
			{Q.}~\bibnamefont {Huang}},\ and\ \bibinfo {author} {\bibfnamefont
			{P.}~\bibnamefont {Dai}},\ }\bibfield  {title} {\bibinfo {title} {Topological
			spin excitations in honeycomb ferromagnet {${\mathrm{CrI}}_{3}$}},\ }\href
	{https://doi.org/10.1103/PhysRevX.8.041028} {\bibfield  {journal} {\bibinfo
			{journal} {Phys. Rev. X}\ }\textbf {\bibinfo {volume} {8}},\ \bibinfo {pages}
		{041028} (\bibinfo {year} {2018})}\BibitemShut {NoStop}%
	\bibitem [{\citenamefont {G{\"o}ser}\ \emph {et~al.}(1990)\citenamefont
		{G{\"o}ser}, \citenamefont {Paul},\ and\ \citenamefont
		{Kahle}}]{goser1990magnetic}%
	\BibitemOpen
	\bibfield  {author} {\bibinfo {author} {\bibfnamefont {O.}~\bibnamefont
			{G{\"o}ser}}, \bibinfo {author} {\bibfnamefont {W.}~\bibnamefont {Paul}},\
		and\ \bibinfo {author} {\bibfnamefont {H.}~\bibnamefont {Kahle}},\ }\bibfield
	{title} {\bibinfo {title} {Magnetic properties of {CrSBr}},\ }\href@noop {}
	{\bibfield  {journal} {\bibinfo  {journal} {J. Magn. Magn. Mater.}\ }\textbf
		{\bibinfo {volume} {92}},\ \bibinfo {pages} {129} (\bibinfo {year}
		{1990})}\BibitemShut {NoStop}%
	\bibitem [{\citenamefont {Telford}\ \emph {et~al.}(2020)\citenamefont
		{Telford}, \citenamefont {Dismukes}, \citenamefont {Lee}, \citenamefont
		{Cheng}, \citenamefont {Wieteska}, \citenamefont {Bartholomew}, \citenamefont
		{Chen}, \citenamefont {Xu}, \citenamefont {Pasupathy}, \citenamefont {Zhu}
		\emph {et~al.}}]{telford2020layered}%
	\BibitemOpen
	\bibfield  {author} {\bibinfo {author} {\bibfnamefont {E.~J.}\ \bibnamefont
			{Telford}}, \bibinfo {author} {\bibfnamefont {A.~H.}\ \bibnamefont
			{Dismukes}}, \bibinfo {author} {\bibfnamefont {K.}~\bibnamefont {Lee}},
		\bibinfo {author} {\bibfnamefont {M.}~\bibnamefont {Cheng}}, \bibinfo
		{author} {\bibfnamefont {A.}~\bibnamefont {Wieteska}}, \bibinfo {author}
		{\bibfnamefont {A.~K.}\ \bibnamefont {Bartholomew}}, \bibinfo {author}
		{\bibfnamefont {Y.-S.}\ \bibnamefont {Chen}}, \bibinfo {author}
		{\bibfnamefont {X.}~\bibnamefont {Xu}}, \bibinfo {author} {\bibfnamefont
			{A.~N.}\ \bibnamefont {Pasupathy}}, \bibinfo {author} {\bibfnamefont
			{X.}~\bibnamefont {Zhu}}, \emph {et~al.},\ }\bibfield  {title} {\bibinfo
		{title} {Layered antiferromagnetism induces large negative magnetoresistance
			in the van der {Waals} semiconductor {CrSBr}},\ }\href
	{https://doi.org/10.1002/adma.202003240} {\bibfield  {journal} {\bibinfo
			{journal} {Advanced Materials}\ }\textbf {\bibinfo {volume} {32}},\ \bibinfo
		{pages} {2003240} (\bibinfo {year} {2020})}\BibitemShut {NoStop}%
	\bibitem [{\citenamefont {L{\'o}pez-Paz}\ \emph {et~al.}(2022)\citenamefont
		{L{\'o}pez-Paz}, \citenamefont {Guguchia}, \citenamefont {Pomjakushin},
		\citenamefont {Witteveen}, \citenamefont {Cervellino}, \citenamefont
		{Luetkens}, \citenamefont {Casati}, \citenamefont {Morpurgo},\ and\
		\citenamefont {von Rohr}}]{lopez2022dynamic}%
	\BibitemOpen
	\bibfield  {author} {\bibinfo {author} {\bibfnamefont {S.~A.}\ \bibnamefont
			{L{\'o}pez-Paz}}, \bibinfo {author} {\bibfnamefont {Z.}~\bibnamefont
			{Guguchia}}, \bibinfo {author} {\bibfnamefont {V.~Y.}\ \bibnamefont
			{Pomjakushin}}, \bibinfo {author} {\bibfnamefont {C.}~\bibnamefont
			{Witteveen}}, \bibinfo {author} {\bibfnamefont {A.}~\bibnamefont
			{Cervellino}}, \bibinfo {author} {\bibfnamefont {H.}~\bibnamefont
			{Luetkens}}, \bibinfo {author} {\bibfnamefont {N.}~\bibnamefont {Casati}},
		\bibinfo {author} {\bibfnamefont {A.~F.}\ \bibnamefont {Morpurgo}},\ and\
		\bibinfo {author} {\bibfnamefont {F.~O.}\ \bibnamefont {von Rohr}},\
	}\bibfield  {title} {\bibinfo {title} {Dynamic magnetic crossover at the
			origin of the hidden-order in van der waals antiferromagnet {CrSBr}},\ }\href
	{https://doi.org/10.48550/arXiv.2203.11785} {\bibfield  {journal} {\bibinfo
			{journal} {arXiv preprint arXiv:2203.11785}\ } (\bibinfo {year}
		{2022})}\BibitemShut {NoStop}%
	\bibitem [{\citenamefont {Lee}\ \emph {et~al.}(2021)\citenamefont {Lee},
		\citenamefont {Dismukes}, \citenamefont {Telford}, \citenamefont {Wiscons},
		\citenamefont {Wang}, \citenamefont {Xu}, \citenamefont {Nuckolls},
		\citenamefont {Dean}, \citenamefont {Roy},\ and\ \citenamefont
		{Zhu}}]{lee2021magnetic}%
	\BibitemOpen
	\bibfield  {author} {\bibinfo {author} {\bibfnamefont {K.}~\bibnamefont
			{Lee}}, \bibinfo {author} {\bibfnamefont {A.~H.}\ \bibnamefont {Dismukes}},
		\bibinfo {author} {\bibfnamefont {E.~J.}\ \bibnamefont {Telford}}, \bibinfo
		{author} {\bibfnamefont {R.~A.}\ \bibnamefont {Wiscons}}, \bibinfo {author}
		{\bibfnamefont {J.}~\bibnamefont {Wang}}, \bibinfo {author} {\bibfnamefont
			{X.}~\bibnamefont {Xu}}, \bibinfo {author} {\bibfnamefont {C.}~\bibnamefont
			{Nuckolls}}, \bibinfo {author} {\bibfnamefont {C.~R.}\ \bibnamefont {Dean}},
		\bibinfo {author} {\bibfnamefont {X.}~\bibnamefont {Roy}},\ and\ \bibinfo
		{author} {\bibfnamefont {X.}~\bibnamefont {Zhu}},\ }\bibfield  {title}
	{\bibinfo {title} {Magnetic order and symmetry in the 2d semiconductor
			{CrSBr}},\ }\href {https://doi.org/10.1021/acs.nanolett.1c00219} {\bibfield
		{journal} {\bibinfo  {journal} {Nano Letters}\ }\textbf {\bibinfo {volume}
			{21}},\ \bibinfo {pages} {3511} (\bibinfo {year} {2021})}\BibitemShut
	{NoStop}%
	\bibitem [{\citenamefont {Wang}\ \emph
		{et~al.}(2020{\natexlab{a}})\citenamefont {Wang}, \citenamefont {Qi},\ and\
		\citenamefont {Qian}}]{qi2018electrically}%
	\BibitemOpen
	\bibfield  {author} {\bibinfo {author} {\bibfnamefont {H.}~\bibnamefont
			{Wang}}, \bibinfo {author} {\bibfnamefont {J.}~\bibnamefont {Qi}},\ and\
		\bibinfo {author} {\bibfnamefont {X.}~\bibnamefont {Qian}},\ }\bibfield
	{title} {\bibinfo {title} {Electrically tunable high curie temperature
			two-dimensional ferromagnetism in van der {Waals} layered crystals},\ }\href
	{https://doi.org/10.1063/5.0014865} {\bibfield  {journal} {\bibinfo
			{journal} {Applied Physics Letters}\ }\textbf {\bibinfo {volume} {117}},\
		\bibinfo {pages} {083102} (\bibinfo {year} {2020}{\natexlab{a}})}\BibitemShut
	{NoStop}%
	\bibitem [{\citenamefont {Telford}\ \emph {et~al.}(2021)\citenamefont
		{Telford}, \citenamefont {Dismukes}, \citenamefont {Dudley}, \citenamefont
		{Wiscons}, \citenamefont {Lee}, \citenamefont {Yu}, \citenamefont {Shabani},
		\citenamefont {Scheie}, \citenamefont {Watanabe}, \citenamefont {Taniguchi}
		\emph {et~al.}}]{telford2021hidden}%
	\BibitemOpen
	\bibfield  {author} {\bibinfo {author} {\bibfnamefont {E.~J.}\ \bibnamefont
			{Telford}}, \bibinfo {author} {\bibfnamefont {A.~H.}\ \bibnamefont
			{Dismukes}}, \bibinfo {author} {\bibfnamefont {R.~L.}\ \bibnamefont
			{Dudley}}, \bibinfo {author} {\bibfnamefont {R.~A.}\ \bibnamefont {Wiscons}},
		\bibinfo {author} {\bibfnamefont {K.}~\bibnamefont {Lee}}, \bibinfo {author}
		{\bibfnamefont {J.}~\bibnamefont {Yu}}, \bibinfo {author} {\bibfnamefont
			{S.}~\bibnamefont {Shabani}}, \bibinfo {author} {\bibfnamefont
			{A.}~\bibnamefont {Scheie}}, \bibinfo {author} {\bibfnamefont
			{K.}~\bibnamefont {Watanabe}}, \bibinfo {author} {\bibfnamefont
			{T.}~\bibnamefont {Taniguchi}}, \emph {et~al.},\ }\bibfield  {title}
	{\bibinfo {title} {Hidden low-temperature magnetic order revealed through
			magnetotransport in monolayer {CrSBr}},\ }\href
	{https://arxiv.org/abs/2106.08471} {\bibfield  {journal} {\bibinfo  {journal}
			{arXiv preprint arXiv:2106.08471}\ } (\bibinfo {year} {2021})}\BibitemShut
	{NoStop}%
	\bibitem [{\citenamefont {Wilson}\ \emph {et~al.}(2021)\citenamefont {Wilson},
		\citenamefont {Lee}, \citenamefont {Cenker}, \citenamefont {Xie},
		\citenamefont {Dismukes}, \citenamefont {Telford}, \citenamefont {Fonseca},
		\citenamefont {Sivakumar}, \citenamefont {Dean}, \citenamefont {Cao},
		\citenamefont {Roy}, \citenamefont {Xu},\ and\ \citenamefont
		{Zhu}}]{Wilson2021}%
	\BibitemOpen
	\bibfield  {author} {\bibinfo {author} {\bibfnamefont {N.~P.}\ \bibnamefont
			{Wilson}}, \bibinfo {author} {\bibfnamefont {K.}~\bibnamefont {Lee}},
		\bibinfo {author} {\bibfnamefont {J.}~\bibnamefont {Cenker}}, \bibinfo
		{author} {\bibfnamefont {K.}~\bibnamefont {Xie}}, \bibinfo {author}
		{\bibfnamefont {A.~H.}\ \bibnamefont {Dismukes}}, \bibinfo {author}
		{\bibfnamefont {E.~J.}\ \bibnamefont {Telford}}, \bibinfo {author}
		{\bibfnamefont {J.}~\bibnamefont {Fonseca}}, \bibinfo {author} {\bibfnamefont
			{S.}~\bibnamefont {Sivakumar}}, \bibinfo {author} {\bibfnamefont
			{C.}~\bibnamefont {Dean}}, \bibinfo {author} {\bibfnamefont {T.}~\bibnamefont
			{Cao}}, \bibinfo {author} {\bibfnamefont {X.}~\bibnamefont {Roy}}, \bibinfo
		{author} {\bibfnamefont {X.}~\bibnamefont {Xu}},\ and\ \bibinfo {author}
		{\bibfnamefont {X.}~\bibnamefont {Zhu}},\ }\bibfield  {title} {\bibinfo
		{title} {Interlayer electronic coupling on demand in a 2d magnetic
			semiconductor},\ }\href {https://doi.org/10.1038/s41563-021-01070-8}
	{\bibfield  {journal} {\bibinfo  {journal} {Nature Materials}\ }\textbf
		{\bibinfo {volume} {20}},\ \bibinfo {pages} {1657} (\bibinfo {year}
		{2021})}\BibitemShut {NoStop}%
	\bibitem [{\citenamefont {Ghiasi}\ \emph {et~al.}(2021)\citenamefont {Ghiasi},
		\citenamefont {Kaverzin}, \citenamefont {Dismukes}, \citenamefont {de~Wal},
		\citenamefont {Roy},\ and\ \citenamefont {van Wees}}]{Ghiasi2021}%
	\BibitemOpen
	\bibfield  {author} {\bibinfo {author} {\bibfnamefont {T.~S.}\ \bibnamefont
			{Ghiasi}}, \bibinfo {author} {\bibfnamefont {A.~A.}\ \bibnamefont
			{Kaverzin}}, \bibinfo {author} {\bibfnamefont {A.~H.}\ \bibnamefont
			{Dismukes}}, \bibinfo {author} {\bibfnamefont {D.~K.}\ \bibnamefont
			{de~Wal}}, \bibinfo {author} {\bibfnamefont {X.}~\bibnamefont {Roy}},\ and\
		\bibinfo {author} {\bibfnamefont {B.~J.}\ \bibnamefont {van Wees}},\
	}\bibfield  {title} {\bibinfo {title} {Electrical and thermal generation of
			spin currents by magnetic bilayer graphene},\ }\href
	{https://doi.org/10.1038/s41565-021-00887-3} {\bibfield  {journal} {\bibinfo
			{journal} {Nature Nanotechnology}\ }\textbf {\bibinfo {volume} {16}},\
		\bibinfo {pages} {788} (\bibinfo {year} {2021})}\BibitemShut {NoStop}%
	\bibitem [{\citenamefont {Chaikin}\ \emph {et~al.}(1995)\citenamefont
		{Chaikin}, \citenamefont {Lubensky},\ and\ \citenamefont
		{Witten}}]{chaikin1995principles}%
	\BibitemOpen
	\bibfield  {author} {\bibinfo {author} {\bibfnamefont {P.~M.}\ \bibnamefont
			{Chaikin}}, \bibinfo {author} {\bibfnamefont {T.~C.}\ \bibnamefont
			{Lubensky}},\ and\ \bibinfo {author} {\bibfnamefont {T.~A.}\ \bibnamefont
			{Witten}},\ }\href@noop {} {\emph {\bibinfo {title} {Principles of condensed
				matter physics}}},\ Vol.~\bibinfo {volume} {10}\ (\bibinfo  {publisher}
	{Cambridge university press Cambridge},\ \bibinfo {year} {1995})\BibitemShut
	{NoStop}%
	\bibitem [{\citenamefont {Bramwell}\ and\ \citenamefont
		{Holdsworth}(1993)}]{Bramwell_1993}%
	\BibitemOpen
	\bibfield  {author} {\bibinfo {author} {\bibfnamefont {S.~T.}\ \bibnamefont
			{Bramwell}}\ and\ \bibinfo {author} {\bibfnamefont {P.~C.~W.}\ \bibnamefont
			{Holdsworth}},\ }\bibfield  {title} {\bibinfo {title} {Magnetization and
			universal sub-critical behaviour in two-dimensional {XY} magnets},\ }\href
	{https://doi.org/10.1088/0953-8984/5/4/004} {\bibfield  {journal} {\bibinfo
			{journal} {Journal of Physics: Condensed Matter}\ }\textbf {\bibinfo {volume}
			{5}},\ \bibinfo {pages} {L53} (\bibinfo {year} {1993})}\BibitemShut {NoStop}%
	\bibitem [{\citenamefont {Wang}\ \emph
		{et~al.}(2020{\natexlab{b}})\citenamefont {Wang}, \citenamefont {Qi},\ and\
		\citenamefont {Qian}}]{Wang_2020_Electrically}%
	\BibitemOpen
	\bibfield  {author} {\bibinfo {author} {\bibfnamefont {H.}~\bibnamefont
			{Wang}}, \bibinfo {author} {\bibfnamefont {J.}~\bibnamefont {Qi}},\ and\
		\bibinfo {author} {\bibfnamefont {X.}~\bibnamefont {Qian}},\ }\href
	{https://doi.org/10.1063/5.0014865} {\bibfield  {journal} {\bibinfo
			{journal} {Applied Physics Letters}\ }\textbf {\bibinfo {volume} {117}},\
		\bibinfo {pages} {083102} (\bibinfo {year} {2020}{\natexlab{b}})}\BibitemShut
	{NoStop}%
	\bibitem [{\citenamefont {Bae}\ \emph {et~al.}(2022)\citenamefont {Bae},
		\citenamefont {Wang}, \citenamefont {Xu}, \citenamefont {Chica},
		\citenamefont {Diederich}, \citenamefont {Cenker}, \citenamefont {Ziebel},
		\citenamefont {Bai}, \citenamefont {Ren}, \citenamefont {Dean} \emph
		{et~al.}}]{bae2022exciton}%
	\BibitemOpen
	\bibfield  {author} {\bibinfo {author} {\bibfnamefont {Y.~J.}\ \bibnamefont
			{Bae}}, \bibinfo {author} {\bibfnamefont {J.}~\bibnamefont {Wang}}, \bibinfo
		{author} {\bibfnamefont {J.}~\bibnamefont {Xu}}, \bibinfo {author}
		{\bibfnamefont {D.~G.}\ \bibnamefont {Chica}}, \bibinfo {author}
		{\bibfnamefont {G.~M.}\ \bibnamefont {Diederich}}, \bibinfo {author}
		{\bibfnamefont {J.}~\bibnamefont {Cenker}}, \bibinfo {author} {\bibfnamefont
			{M.~E.}\ \bibnamefont {Ziebel}}, \bibinfo {author} {\bibfnamefont
			{Y.}~\bibnamefont {Bai}}, \bibinfo {author} {\bibfnamefont {H.}~\bibnamefont
			{Ren}}, \bibinfo {author} {\bibfnamefont {C.~R.}\ \bibnamefont {Dean}}, \emph
		{et~al.},\ }\bibfield  {title} {\bibinfo {title} {Exciton-coupled coherent
			antiferromagnetic magnons in a 2d semiconductor},\ }\href
	{https://doi.org/10.48550/arXiv.2201.13197} {\bibfield  {journal} {\bibinfo
			{journal} {arXiv preprint arXiv:2201.13197}\ } (\bibinfo {year}
		{2022})}\BibitemShut {NoStop}%
	\bibitem [{Sup()}]{SuppMat}%
	\BibitemOpen
	\href@noop {} {}\bibinfo {note} {See Supplemental Material at [URL will be
		inserted by publisher] for more details of the experiments and
		calculations.}\BibitemShut {Stop}%
	\bibitem [{\citenamefont {Yang}\ \emph {et~al.}(2021)\citenamefont {Yang},
		\citenamefont {Wang}, \citenamefont {Liu}, \citenamefont {Lu},\ and\
		\citenamefont {Wu}}]{Yang_2021}%
	\BibitemOpen
	\bibfield  {author} {\bibinfo {author} {\bibfnamefont {K.}~\bibnamefont
			{Yang}}, \bibinfo {author} {\bibfnamefont {G.}~\bibnamefont {Wang}}, \bibinfo
		{author} {\bibfnamefont {L.}~\bibnamefont {Liu}}, \bibinfo {author}
		{\bibfnamefont {D.}~\bibnamefont {Lu}},\ and\ \bibinfo {author}
		{\bibfnamefont {H.}~\bibnamefont {Wu}},\ }\bibfield  {title} {\bibinfo
		{title} {Triaxial magnetic anisotropy in the two-dimensional ferromagnetic
			semiconductor {CrSBr}},\ }\href {https://doi.org/10.1103/PhysRevB.104.144416}
	{\bibfield  {journal} {\bibinfo  {journal} {Phys. Rev. B}\ }\textbf {\bibinfo
			{volume} {104}},\ \bibinfo {pages} {144416} (\bibinfo {year}
		{2021})}\BibitemShut {NoStop}%
	\bibitem [{\citenamefont {Jensen}\ and\ \citenamefont
		{Mackintosh}(1991)}]{Jensen+Mackintosh}%
	\BibitemOpen
	\bibfield  {author} {\bibinfo {author} {\bibfnamefont {J.}~\bibnamefont
			{Jensen}}\ and\ \bibinfo {author} {\bibfnamefont {A.}~\bibnamefont
			{Mackintosh}},\ }\href@noop {} {\emph {\bibinfo {title} {Rare Earth
				Magnetism, Structures and Excitations}}}\ (\bibinfo  {publisher} {Clarendon
		Press},\ \bibinfo {address} {Oxford, UK},\ \bibinfo {year}
	{1991})\BibitemShut {NoStop}%
	\bibitem [{\citenamefont {Virtanen}\ \emph {et~al.}(2020)\citenamefont
		{Virtanen}, \citenamefont {Gommers}, \citenamefont {Oliphant}, \citenamefont
		{Haberland}, \citenamefont {Reddy}, \citenamefont {Cournapeau}, \citenamefont
		{Burovski}, \citenamefont {Peterson}, \citenamefont {Weckesser},
		\citenamefont {Bright} \emph {et~al.}}]{virtanen2020scipy}%
	\BibitemOpen
	\bibfield  {author} {\bibinfo {author} {\bibfnamefont {P.}~\bibnamefont
			{Virtanen}}, \bibinfo {author} {\bibfnamefont {R.}~\bibnamefont {Gommers}},
		\bibinfo {author} {\bibfnamefont {T.~E.}\ \bibnamefont {Oliphant}}, \bibinfo
		{author} {\bibfnamefont {M.}~\bibnamefont {Haberland}}, \bibinfo {author}
		{\bibfnamefont {T.}~\bibnamefont {Reddy}}, \bibinfo {author} {\bibfnamefont
			{D.}~\bibnamefont {Cournapeau}}, \bibinfo {author} {\bibfnamefont
			{E.}~\bibnamefont {Burovski}}, \bibinfo {author} {\bibfnamefont
			{P.}~\bibnamefont {Peterson}}, \bibinfo {author} {\bibfnamefont
			{W.}~\bibnamefont {Weckesser}}, \bibinfo {author} {\bibfnamefont
			{J.}~\bibnamefont {Bright}}, \emph {et~al.},\ }\bibfield  {title} {\bibinfo
		{title} {Scipy 1.0: fundamental algorithms for scientific computing in
			python},\ }\href {https://doi.org/10.1038/s41592-019-0686-2} {\bibfield
		{journal} {\bibinfo  {journal} {Nature Methods}\ }\textbf {\bibinfo {volume}
			{17}},\ \bibinfo {pages} {261} (\bibinfo {year} {2020})}\BibitemShut
	{NoStop}%
	\bibitem [{\citenamefont {Press}\ \emph {et~al.}(2007)\citenamefont {Press},
		\citenamefont {Teukolsky}, \citenamefont {Vetterling},\ and\ \citenamefont
		{Flannery}}]{NumericalRecipes}%
	\BibitemOpen
	\bibfield  {author} {\bibinfo {author} {\bibfnamefont {W.~H.}\ \bibnamefont
			{Press}}, \bibinfo {author} {\bibfnamefont {S.~A.}\ \bibnamefont
			{Teukolsky}}, \bibinfo {author} {\bibfnamefont {W.~T.}\ \bibnamefont
			{Vetterling}},\ and\ \bibinfo {author} {\bibfnamefont {B.~P.}\ \bibnamefont
			{Flannery}},\ }\href@noop {} {\emph {\bibinfo {title} {Numerical recipes 3rd
				edition: The art of scientific computing}}}\ (\bibinfo  {publisher}
	{Cambridge university press},\ \bibinfo {year} {2007})\BibitemShut {NoStop}%
	\bibitem [{\citenamefont {Toth}\ and\ \citenamefont {Lake}(2015)}]{SpinW}%
	\BibitemOpen
	\bibfield  {author} {\bibinfo {author} {\bibfnamefont {S.}~\bibnamefont
			{Toth}}\ and\ \bibinfo {author} {\bibfnamefont {B.}~\bibnamefont {Lake}},\
	}\bibfield  {title} {\bibinfo {title} {Linear spin wave theory for single-q
			incommensurate magnetic structures},\ }\href
	{https://doi.org/10.1088/0953-8984/27/16/166002} {\bibfield  {journal}
		{\bibinfo  {journal} {Journal of Physics: Condensed Matter}\ }\textbf
		{\bibinfo {volume} {27}},\ \bibinfo {pages} {166002} (\bibinfo {year}
		{2015})}\BibitemShut {NoStop}%
	\bibitem [{\citenamefont {Moriya}(1960)}]{Moriya_1960}%
	\BibitemOpen
	\bibfield  {author} {\bibinfo {author} {\bibfnamefont {T.}~\bibnamefont
			{Moriya}},\ }\bibfield  {title} {\bibinfo {title} {Anisotropic superexchange
			interaction and weak ferromagnetism},\ }\href
	{https://doi.org/10.1103/PhysRev.120.91} {\bibfield  {journal} {\bibinfo
			{journal} {Phys. Rev.}\ }\textbf {\bibinfo {volume} {120}},\ \bibinfo {pages}
		{91} (\bibinfo {year} {1960})}\BibitemShut {NoStop}%
	\bibitem [{\citenamefont {Wu}\ \emph {et~al.}(2022)\citenamefont {Wu},
		\citenamefont {Gutierrez-Lezama}, \citenamefont {L{\'o}pez-Paz},
		\citenamefont {Gibertini}, \citenamefont {Watanabe}, \citenamefont
		{Taniguchi}, \citenamefont {von Rohr}, \citenamefont {Ubrig},\ and\
		\citenamefont {Morpurgo}}]{Wu_2022_quasi1d}%
	\BibitemOpen
	\bibfield  {author} {\bibinfo {author} {\bibfnamefont {F.}~\bibnamefont
			{Wu}}, \bibinfo {author} {\bibfnamefont {I.}~\bibnamefont
			{Gutierrez-Lezama}}, \bibinfo {author} {\bibfnamefont {S.~A.}\ \bibnamefont
			{L{\'o}pez-Paz}}, \bibinfo {author} {\bibfnamefont {M.}~\bibnamefont
			{Gibertini}}, \bibinfo {author} {\bibfnamefont {K.}~\bibnamefont {Watanabe}},
		\bibinfo {author} {\bibfnamefont {T.}~\bibnamefont {Taniguchi}}, \bibinfo
		{author} {\bibfnamefont {F.~O.}\ \bibnamefont {von Rohr}}, \bibinfo {author}
		{\bibfnamefont {N.}~\bibnamefont {Ubrig}},\ and\ \bibinfo {author}
		{\bibfnamefont {A.~F.}\ \bibnamefont {Morpurgo}},\ }\bibfield  {title}
	{\bibinfo {title} {Quasi-{1D} electronic transport in a {2D} magnetic
			semiconductor},\ }\href
	{https://doi.org/https://doi.org/10.1002/adma.202109759} {\bibfield
		{journal} {\bibinfo  {journal} {Advanced Materials}\ }\textbf {\bibinfo
			{volume} {n/a}},\ \bibinfo {pages} {2109759} (\bibinfo {year}
		{2022})}\BibitemShut {NoStop}%
	\bibitem [{\citenamefont {Guo}\ \emph {et~al.}(2018)\citenamefont {Guo},
		\citenamefont {Zhang}, \citenamefont {Yuan}, \citenamefont {Wang},\ and\
		\citenamefont {Wang}}]{guo2018chromium}%
	\BibitemOpen
	\bibfield  {author} {\bibinfo {author} {\bibfnamefont {Y.}~\bibnamefont
			{Guo}}, \bibinfo {author} {\bibfnamefont {Y.}~\bibnamefont {Zhang}}, \bibinfo
		{author} {\bibfnamefont {S.}~\bibnamefont {Yuan}}, \bibinfo {author}
		{\bibfnamefont {B.}~\bibnamefont {Wang}},\ and\ \bibinfo {author}
		{\bibfnamefont {J.}~\bibnamefont {Wang}},\ }\bibfield  {title} {\bibinfo
		{title} {Chromium sulfide halide monolayers: intrinsic ferromagnetic
			semiconductors with large spin polarization and high carrier mobility},\
	}\href {https://doi.org/10.1039/C8NR06368K} {\bibfield  {journal} {\bibinfo
			{journal} {Nanoscale}\ }\textbf {\bibinfo {volume} {10}},\ \bibinfo {pages}
		{18036} (\bibinfo {year} {2018})}\BibitemShut {NoStop}%
	\bibitem [{\citenamefont {Boix-Constant}\ \emph {et~al.}(2022)\citenamefont
		{Boix-Constant}, \citenamefont {Ma{\~n}as-Valero}, \citenamefont {Ruiz},
		\citenamefont {Rybakov}, \citenamefont {Konieczny}, \citenamefont {Pillet},
		\citenamefont {Baldov{\'\i}},\ and\ \citenamefont
		{Coronado}}]{Boix_2022_Probing}%
	\BibitemOpen
	\bibfield  {author} {\bibinfo {author} {\bibfnamefont {C.}~\bibnamefont
			{Boix-Constant}}, \bibinfo {author} {\bibfnamefont {S.}~\bibnamefont
			{Ma{\~n}as-Valero}}, \bibinfo {author} {\bibfnamefont {A.~M.}\ \bibnamefont
			{Ruiz}}, \bibinfo {author} {\bibfnamefont {A.}~\bibnamefont {Rybakov}},
		\bibinfo {author} {\bibfnamefont {K.~A.}\ \bibnamefont {Konieczny}}, \bibinfo
		{author} {\bibfnamefont {S.}~\bibnamefont {Pillet}}, \bibinfo {author}
		{\bibfnamefont {J.~J.}\ \bibnamefont {Baldov{\'\i}}},\ and\ \bibinfo {author}
		{\bibfnamefont {E.}~\bibnamefont {Coronado}},\ }\bibfield  {title} {\bibinfo
		{title} {Probing the spin dimensionality in single-layer {CrSBr} van der
			{Waals} heterostructures by magneto-transport measurements},\ }\href
	{https://arxiv.org/abs/2204.04095} {\bibfield  {journal} {\bibinfo  {journal}
			{arXiv preprint arXiv:2204.04095}\ } (\bibinfo {year} {2022})}\BibitemShut
	{NoStop}%
	\bibitem [{\citenamefont {Wang}\ \emph {et~al.}(2018)\citenamefont {Wang},
		\citenamefont {Zhang},\ and\ \citenamefont {Wang}}]{Wang_2018_topological}%
	\BibitemOpen
	\bibfield  {author} {\bibinfo {author} {\bibfnamefont {X.~S.}\ \bibnamefont
			{Wang}}, \bibinfo {author} {\bibfnamefont {H.~W.}\ \bibnamefont {Zhang}},\
		and\ \bibinfo {author} {\bibfnamefont {X.~R.}\ \bibnamefont {Wang}},\
	}\bibfield  {title} {\bibinfo {title} {Topological magnonics: A paradigm for
			spin-wave manipulation and device design},\ }\href
	{https://doi.org/10.1103/PhysRevApplied.9.024029} {\bibfield  {journal}
		{\bibinfo  {journal} {Phys. Rev. Applied}\ }\textbf {\bibinfo {volume} {9}},\
		\bibinfo {pages} {024029} (\bibinfo {year} {2018})}\BibitemShut {NoStop}%
	\bibitem [{\citenamefont {Barman}\ \emph {et~al.}(2021)\citenamefont {Barman},
		\citenamefont {Gubbiotti}, \citenamefont {Ladak}, \citenamefont {Adeyeye},
		\citenamefont {Krawczyk}, \citenamefont {Grafe}, \citenamefont {Adelmann},
		\citenamefont {Cotofana}, \citenamefont {Naeemi}, \citenamefont {Vasyuchka},
		\citenamefont {Hillebrands}, \citenamefont {Nikitov}, \citenamefont {Yu},
		\citenamefont {Grundler}, \citenamefont {Sadovnikov}, \citenamefont
		{Grachev}, \citenamefont {Sheshukova}, \citenamefont {Duquesne},
		\citenamefont {Marangolo}, \citenamefont {Csaba}, \citenamefont {Porod},
		\citenamefont {Demidov}, \citenamefont {Urazhdin}, \citenamefont
		{Demokritov}, \citenamefont {Albisetti}, \citenamefont {Petti}, \citenamefont
		{Bertacco}, \citenamefont {Schultheiss}, \citenamefont {Kruglyak},
		\citenamefont {Poimanov}, \citenamefont {Sahoo}, \citenamefont {Sinha},
		\citenamefont {Yang}, \citenamefont {Munzenberg}, \citenamefont {Moriyama},
		\citenamefont {Mizukami}, \citenamefont {Landeros}, \citenamefont {Gallardo},
		\citenamefont {Carlotti}, \citenamefont {Kim}, \citenamefont {Stamps},
		\citenamefont {Camley}, \citenamefont {Rana}, \citenamefont {Otani},
		\citenamefont {Yu}, \citenamefont {Yu}, \citenamefont {Bauer}, \citenamefont
		{Back}, \citenamefont {Uhrig}, \citenamefont {Dobrovolskiy}, \citenamefont
		{Budinska}, \citenamefont {Qin}, \citenamefont {van Dijken}, \citenamefont
		{Chumak}, \citenamefont {Khitun}, \citenamefont {Nikonov}, \citenamefont
		{Young}, \citenamefont {Zingsem},\ and\ \citenamefont
		{Winklhofer}}]{Barman_2021}%
	\BibitemOpen
	\bibfield  {author} {\bibinfo {author} {\bibfnamefont {A.}~\bibnamefont
			{Barman}}, \bibinfo {author} {\bibfnamefont {G.}~\bibnamefont {Gubbiotti}},
		\bibinfo {author} {\bibfnamefont {S.}~\bibnamefont {Ladak}}, \bibinfo
		{author} {\bibfnamefont {A.~O.}\ \bibnamefont {Adeyeye}}, \bibinfo {author}
		{\bibfnamefont {M.}~\bibnamefont {Krawczyk}}, \bibinfo {author}
		{\bibfnamefont {J.}~\bibnamefont {Grafe}}, \bibinfo {author} {\bibfnamefont
			{C.}~\bibnamefont {Adelmann}}, \bibinfo {author} {\bibfnamefont
			{S.}~\bibnamefont {Cotofana}}, \bibinfo {author} {\bibfnamefont
			{A.}~\bibnamefont {Naeemi}}, \bibinfo {author} {\bibfnamefont {V.~I.}\
			\bibnamefont {Vasyuchka}}, \bibinfo {author} {\bibfnamefont {B.}~\bibnamefont
			{Hillebrands}}, \bibinfo {author} {\bibfnamefont {S.~A.}\ \bibnamefont
			{Nikitov}}, \bibinfo {author} {\bibfnamefont {H.}~\bibnamefont {Yu}},
		\bibinfo {author} {\bibfnamefont {D.}~\bibnamefont {Grundler}}, \bibinfo
		{author} {\bibfnamefont {A.~V.}\ \bibnamefont {Sadovnikov}}, \bibinfo
		{author} {\bibfnamefont {A.~A.}\ \bibnamefont {Grachev}}, \bibinfo {author}
		{\bibfnamefont {S.~E.}\ \bibnamefont {Sheshukova}}, \bibinfo {author}
		{\bibfnamefont {J.-Y.}\ \bibnamefont {Duquesne}}, \bibinfo {author}
		{\bibfnamefont {M.}~\bibnamefont {Marangolo}}, \bibinfo {author}
		{\bibfnamefont {G.}~\bibnamefont {Csaba}}, \bibinfo {author} {\bibfnamefont
			{W.}~\bibnamefont {Porod}}, \bibinfo {author} {\bibfnamefont {V.~E.}\
			\bibnamefont {Demidov}}, \bibinfo {author} {\bibfnamefont {S.}~\bibnamefont
			{Urazhdin}}, \bibinfo {author} {\bibfnamefont {S.~O.}\ \bibnamefont
			{Demokritov}}, \bibinfo {author} {\bibfnamefont {E.}~\bibnamefont
			{Albisetti}}, \bibinfo {author} {\bibfnamefont {D.}~\bibnamefont {Petti}},
		\bibinfo {author} {\bibfnamefont {R.}~\bibnamefont {Bertacco}}, \bibinfo
		{author} {\bibfnamefont {H.}~\bibnamefont {Schultheiss}}, \bibinfo {author}
		{\bibfnamefont {V.~V.}\ \bibnamefont {Kruglyak}}, \bibinfo {author}
		{\bibfnamefont {V.~D.}\ \bibnamefont {Poimanov}}, \bibinfo {author}
		{\bibfnamefont {S.}~\bibnamefont {Sahoo}}, \bibinfo {author} {\bibfnamefont
			{J.}~\bibnamefont {Sinha}}, \bibinfo {author} {\bibfnamefont
			{H.}~\bibnamefont {Yang}}, \bibinfo {author} {\bibfnamefont {M.}~\bibnamefont
			{Munzenberg}}, \bibinfo {author} {\bibfnamefont {T.}~\bibnamefont
			{Moriyama}}, \bibinfo {author} {\bibfnamefont {S.}~\bibnamefont {Mizukami}},
		\bibinfo {author} {\bibfnamefont {P.}~\bibnamefont {Landeros}}, \bibinfo
		{author} {\bibfnamefont {R.~A.}\ \bibnamefont {Gallardo}}, \bibinfo {author}
		{\bibfnamefont {G.}~\bibnamefont {Carlotti}}, \bibinfo {author}
		{\bibfnamefont {J.-V.}\ \bibnamefont {Kim}}, \bibinfo {author} {\bibfnamefont
			{R.~L.}\ \bibnamefont {Stamps}}, \bibinfo {author} {\bibfnamefont {R.~E.}\
			\bibnamefont {Camley}}, \bibinfo {author} {\bibfnamefont {B.}~\bibnamefont
			{Rana}}, \bibinfo {author} {\bibfnamefont {Y.}~\bibnamefont {Otani}},
		\bibinfo {author} {\bibfnamefont {W.}~\bibnamefont {Yu}}, \bibinfo {author}
		{\bibfnamefont {T.}~\bibnamefont {Yu}}, \bibinfo {author} {\bibfnamefont
			{G.~E.~W.}\ \bibnamefont {Bauer}}, \bibinfo {author} {\bibfnamefont
			{C.}~\bibnamefont {Back}}, \bibinfo {author} {\bibfnamefont {G.~S.}\
			\bibnamefont {Uhrig}}, \bibinfo {author} {\bibfnamefont {O.~V.}\ \bibnamefont
			{Dobrovolskiy}}, \bibinfo {author} {\bibfnamefont {B.}~\bibnamefont
			{Budinska}}, \bibinfo {author} {\bibfnamefont {H.}~\bibnamefont {Qin}},
		\bibinfo {author} {\bibfnamefont {S.}~\bibnamefont {van Dijken}}, \bibinfo
		{author} {\bibfnamefont {A.~V.}\ \bibnamefont {Chumak}}, \bibinfo {author}
		{\bibfnamefont {A.}~\bibnamefont {Khitun}}, \bibinfo {author} {\bibfnamefont
			{D.~E.}\ \bibnamefont {Nikonov}}, \bibinfo {author} {\bibfnamefont {I.~A.}\
			\bibnamefont {Young}}, \bibinfo {author} {\bibfnamefont {B.~W.}\ \bibnamefont
			{Zingsem}},\ and\ \bibinfo {author} {\bibfnamefont {M.}~\bibnamefont
			{Winklhofer}},\ }\bibfield  {title} {\bibinfo {title} {The 2021 magnonics
			roadmap},\ }\href {https://doi.org/10.1088/1361-648x/abec1a} {\bibfield
		{journal} {\bibinfo  {journal} {Journal of Physics: Condensed Matter}\
		}\textbf {\bibinfo {volume} {33}},\ \bibinfo {pages} {413001} (\bibinfo
		{year} {2021})}\BibitemShut {NoStop}%
	\bibitem [{\citenamefont {McClarty}(2021)}]{McClarty_2021}%
	\BibitemOpen
	\bibfield  {author} {\bibinfo {author} {\bibfnamefont {P.}~\bibnamefont
			{McClarty}},\ }\bibfield  {title} {\bibinfo {title} {Topological magnons: A
			review},\ }\href {https://arxiv.org/abs/2106.01430} {\bibfield  {journal}
		{\bibinfo  {journal} {arXiv preprint arXiv:2106.01430}\ } (\bibinfo {year}
		{2021})}\BibitemShut {NoStop}%
	\bibitem [{\citenamefont {Bihlmayer}\ \emph {et~al.}(2015)\citenamefont
		{Bihlmayer}, \citenamefont {Rader},\ and\ \citenamefont
		{Winkler}}]{Bihlmayer_2015}%
	\BibitemOpen
	\bibfield  {author} {\bibinfo {author} {\bibfnamefont {G.}~\bibnamefont
			{Bihlmayer}}, \bibinfo {author} {\bibfnamefont {O.}~\bibnamefont {Rader}},\
		and\ \bibinfo {author} {\bibfnamefont {R.}~\bibnamefont {Winkler}},\
	}\bibfield  {title} {\bibinfo {title} {Focus on the rashba effect},\ }\href
	{https://doi.org/10.1088/1367-2630/17/5/050202} {\bibfield  {journal}
		{\bibinfo  {journal} {New Journal of Physics}\ }\textbf {\bibinfo {volume}
			{17}},\ \bibinfo {pages} {050202} (\bibinfo {year} {2015})}\BibitemShut
	{NoStop}%
	\bibitem [{\citenamefont {Fan}\ \emph {et~al.}(2014)\citenamefont {Fan},
		\citenamefont {Celik}, \citenamefont {Wu}, \citenamefont {Ni}, \citenamefont
		{Lee}, \citenamefont {Lorenz},\ and\ \citenamefont {Xiao}}]{Fan2014}%
	\BibitemOpen
	\bibfield  {author} {\bibinfo {author} {\bibfnamefont {X.}~\bibnamefont
			{Fan}}, \bibinfo {author} {\bibfnamefont {H.}~\bibnamefont {Celik}}, \bibinfo
		{author} {\bibfnamefont {J.}~\bibnamefont {Wu}}, \bibinfo {author}
		{\bibfnamefont {C.}~\bibnamefont {Ni}}, \bibinfo {author} {\bibfnamefont
			{K.-J.}\ \bibnamefont {Lee}}, \bibinfo {author} {\bibfnamefont {V.~O.}\
			\bibnamefont {Lorenz}},\ and\ \bibinfo {author} {\bibfnamefont {J.~Q.}\
			\bibnamefont {Xiao}},\ }\bibfield  {title} {\bibinfo {title} {Quantifying
			interface and bulk contributions to spin--orbit torque in magnetic
			bilayers},\ }\href {https://doi.org/10.1038/ncomms4042} {\bibfield  {journal}
		{\bibinfo  {journal} {Nature Communications}\ }\textbf {\bibinfo {volume}
			{5}},\ \bibinfo {pages} {3042} (\bibinfo {year} {2014})}\BibitemShut
	{NoStop}%
	\bibitem [{\citenamefont {Avsar}\ \emph {et~al.}(2014)\citenamefont {Avsar},
		\citenamefont {Tan}, \citenamefont {Taychatanapat}, \citenamefont
		{Balakrishnan}, \citenamefont {Koon}, \citenamefont {Yeo}, \citenamefont
		{Lahiri}, \citenamefont {Carvalho}, \citenamefont {Rodin}, \citenamefont
		{O'Farrell} \emph {et~al.}}]{avsar2014spin}%
	\BibitemOpen
	\bibfield  {author} {\bibinfo {author} {\bibfnamefont {A.}~\bibnamefont
			{Avsar}}, \bibinfo {author} {\bibfnamefont {J.~Y.}\ \bibnamefont {Tan}},
		\bibinfo {author} {\bibfnamefont {T.}~\bibnamefont {Taychatanapat}}, \bibinfo
		{author} {\bibfnamefont {J.}~\bibnamefont {Balakrishnan}}, \bibinfo {author}
		{\bibfnamefont {G.}~\bibnamefont {Koon}}, \bibinfo {author} {\bibfnamefont
			{Y.}~\bibnamefont {Yeo}}, \bibinfo {author} {\bibfnamefont {J.}~\bibnamefont
			{Lahiri}}, \bibinfo {author} {\bibfnamefont {A.}~\bibnamefont {Carvalho}},
		\bibinfo {author} {\bibfnamefont {A.}~\bibnamefont {Rodin}}, \bibinfo
		{author} {\bibfnamefont {E.}~\bibnamefont {O'Farrell}}, \emph {et~al.},\
	}\bibfield  {title} {\bibinfo {title} {Spin-orbit proximity effect in
			graphene},\ }\href {https://doi.org/10.1038/ncomms5875} {\bibfield  {journal}
		{\bibinfo  {journal} {Nature communications}\ }\textbf {\bibinfo {volume}
			{5}},\ \bibinfo {pages} {1} (\bibinfo {year} {2014})}\BibitemShut {NoStop}%
	\bibitem [{\citenamefont {Wang}\ \emph {et~al.}(2016)\citenamefont {Wang},
		\citenamefont {Ki}, \citenamefont {Khoo}, \citenamefont {Mauro},
		\citenamefont {Berger}, \citenamefont {Levitov},\ and\ \citenamefont
		{Morpurgo}}]{Wang_2016_graphene}%
	\BibitemOpen
	\bibfield  {author} {\bibinfo {author} {\bibfnamefont {Z.}~\bibnamefont
			{Wang}}, \bibinfo {author} {\bibfnamefont {D.-K.}\ \bibnamefont {Ki}},
		\bibinfo {author} {\bibfnamefont {J.~Y.}\ \bibnamefont {Khoo}}, \bibinfo
		{author} {\bibfnamefont {D.}~\bibnamefont {Mauro}}, \bibinfo {author}
		{\bibfnamefont {H.}~\bibnamefont {Berger}}, \bibinfo {author} {\bibfnamefont
			{L.~S.}\ \bibnamefont {Levitov}},\ and\ \bibinfo {author} {\bibfnamefont
			{A.~F.}\ \bibnamefont {Morpurgo}},\ }\bibfield  {title} {\bibinfo {title}
		{Origin and magnitude of `designer' spin-orbit interaction in graphene on
			semiconducting transition metal dichalcogenides},\ }\href
	{https://doi.org/10.1103/PhysRevX.6.041020} {\bibfield  {journal} {\bibinfo
			{journal} {Phys. Rev. X}\ }\textbf {\bibinfo {volume} {6}},\ \bibinfo {pages}
		{041020} (\bibinfo {year} {2016})}\BibitemShut {NoStop}%
	\bibitem [{\citenamefont {Yang}\ \emph {et~al.}(2020)\citenamefont {Yang},
		\citenamefont {Li}, \citenamefont {Chopdekar}, \citenamefont {Dhall},
		\citenamefont {Turner}, \citenamefont {Carlstrom}, \citenamefont {Ophus},
		\citenamefont {Klewe}, \citenamefont {Shafer}, \citenamefont {N'Diaye},
		\citenamefont {Choi}, \citenamefont {Chen}, \citenamefont {Wu}, \citenamefont
		{Hwang}, \citenamefont {Wang},\ and\ \citenamefont
		{Qiu}}]{Yang_2020_Creation}%
	\BibitemOpen
	\bibfield  {author} {\bibinfo {author} {\bibfnamefont {M.}~\bibnamefont
			{Yang}}, \bibinfo {author} {\bibfnamefont {Q.}~\bibnamefont {Li}}, \bibinfo
		{author} {\bibfnamefont {R.~V.}\ \bibnamefont {Chopdekar}}, \bibinfo {author}
		{\bibfnamefont {R.}~\bibnamefont {Dhall}}, \bibinfo {author} {\bibfnamefont
			{J.}~\bibnamefont {Turner}}, \bibinfo {author} {\bibfnamefont {J.~D.}\
			\bibnamefont {Carlstrom}}, \bibinfo {author} {\bibfnamefont {C.}~\bibnamefont
			{Ophus}}, \bibinfo {author} {\bibfnamefont {C.}~\bibnamefont {Klewe}},
		\bibinfo {author} {\bibfnamefont {P.}~\bibnamefont {Shafer}}, \bibinfo
		{author} {\bibfnamefont {A.~T.}\ \bibnamefont {N'Diaye}}, \bibinfo {author}
		{\bibfnamefont {J.~W.}\ \bibnamefont {Choi}}, \bibinfo {author}
		{\bibfnamefont {G.}~\bibnamefont {Chen}}, \bibinfo {author} {\bibfnamefont
			{Y.~Z.}\ \bibnamefont {Wu}}, \bibinfo {author} {\bibfnamefont
			{C.}~\bibnamefont {Hwang}}, \bibinfo {author} {\bibfnamefont
			{F.}~\bibnamefont {Wang}},\ and\ \bibinfo {author} {\bibfnamefont {Z.~Q.}\
			\bibnamefont {Qiu}},\ }\bibfield  {title} {\bibinfo {title} {Creation of
			skyrmions in van der {Waals} ferromagnet {Fe$_3$GeTe$_2$} on {(Co/Pd)$_n$}
			superlattice},\ }\href {https://doi.org/10.1126/sciadv.abb5157} {\bibfield
		{journal} {\bibinfo  {journal} {Science Advances}\ }\textbf {\bibinfo
			{volume} {6}},\ \bibinfo {pages} {eabb5157} (\bibinfo {year}
		{2020})}\BibitemShut {NoStop}%
	\bibitem [{\citenamefont {Wu}\ \emph {et~al.}(2020)\citenamefont {Wu},
		\citenamefont {Zhang}, \citenamefont {Zhang}, \citenamefont {Wang},
		\citenamefont {Zhu}, \citenamefont {Hu}, \citenamefont {Yin}, \citenamefont
		{Wong}, \citenamefont {Fang}, \citenamefont {Wan} \emph
		{et~al.}}]{wu2020neel}%
	\BibitemOpen
	\bibfield  {author} {\bibinfo {author} {\bibfnamefont {Y.}~\bibnamefont
			{Wu}}, \bibinfo {author} {\bibfnamefont {S.}~\bibnamefont {Zhang}}, \bibinfo
		{author} {\bibfnamefont {J.}~\bibnamefont {Zhang}}, \bibinfo {author}
		{\bibfnamefont {W.}~\bibnamefont {Wang}}, \bibinfo {author} {\bibfnamefont
			{Y.~L.}\ \bibnamefont {Zhu}}, \bibinfo {author} {\bibfnamefont
			{J.}~\bibnamefont {Hu}}, \bibinfo {author} {\bibfnamefont {G.}~\bibnamefont
			{Yin}}, \bibinfo {author} {\bibfnamefont {K.}~\bibnamefont {Wong}}, \bibinfo
		{author} {\bibfnamefont {C.}~\bibnamefont {Fang}}, \bibinfo {author}
		{\bibfnamefont {C.}~\bibnamefont {Wan}}, \emph {et~al.},\ }\bibfield  {title}
	{\bibinfo {title} {N{\'e}el-type skyrmion in {WTe$_2$}/{Fe$_3$GeTe$_2$} van
			der {Waals} heterostructure},\ }\href
	{https://doi.org/10.1038/s41467-020-17566-x} {\bibfield  {journal} {\bibinfo
			{journal} {Nature communications}\ }\textbf {\bibinfo {volume} {11}},\
		\bibinfo {pages} {1} (\bibinfo {year} {2020})}\BibitemShut {NoStop}%
	\bibitem [{\citenamefont {Coates}\ \emph {et~al.}(2018)\citenamefont {Coates},
		\citenamefont {Cao}, \citenamefont {Chakoumakos}, \citenamefont {Frontzek},
		\citenamefont {Hoffmann}, \citenamefont {Kovalevsky}, \citenamefont {Liu},
		\citenamefont {Meilleur}, \citenamefont {dos Santos}, \citenamefont {Myles}
		\emph {et~al.}}]{coates2018suite}%
	\BibitemOpen
	\bibfield  {author} {\bibinfo {author} {\bibfnamefont {L.}~\bibnamefont
			{Coates}}, \bibinfo {author} {\bibfnamefont {H.}~\bibnamefont {Cao}},
		\bibinfo {author} {\bibfnamefont {B.~C.}\ \bibnamefont {Chakoumakos}},
		\bibinfo {author} {\bibfnamefont {M.~D.}\ \bibnamefont {Frontzek}}, \bibinfo
		{author} {\bibfnamefont {C.}~\bibnamefont {Hoffmann}}, \bibinfo {author}
		{\bibfnamefont {A.~Y.}\ \bibnamefont {Kovalevsky}}, \bibinfo {author}
		{\bibfnamefont {Y.}~\bibnamefont {Liu}}, \bibinfo {author} {\bibfnamefont
			{F.}~\bibnamefont {Meilleur}}, \bibinfo {author} {\bibfnamefont {A.~M.}\
			\bibnamefont {dos Santos}}, \bibinfo {author} {\bibfnamefont {D.~A.}\
			\bibnamefont {Myles}}, \emph {et~al.},\ }\bibfield  {title} {\bibinfo {title}
		{A suite-level review of the neutron single-crystal diffraction instruments
			at oak ridge national laboratory},\ }\href
	{https://doi.org/10.1063/1.5030896} {\bibfield  {journal} {\bibinfo
			{journal} {Review of Scientific Instruments}\ }\textbf {\bibinfo {volume}
			{89}},\ \bibinfo {pages} {092802} (\bibinfo {year} {2018})}\BibitemShut
	{NoStop}%
	\bibitem [{\citenamefont {Zikovsky}\ \emph {et~al.}(2011)\citenamefont
		{Zikovsky}, \citenamefont {Peterson}, \citenamefont {Wang}, \citenamefont
		{Frost},\ and\ \citenamefont {Hoffmann}}]{zikovsky2011crystalplan}%
	\BibitemOpen
	\bibfield  {author} {\bibinfo {author} {\bibfnamefont {J.}~\bibnamefont
			{Zikovsky}}, \bibinfo {author} {\bibfnamefont {P.~F.}\ \bibnamefont
			{Peterson}}, \bibinfo {author} {\bibfnamefont {X.~P.}\ \bibnamefont {Wang}},
		\bibinfo {author} {\bibfnamefont {M.}~\bibnamefont {Frost}},\ and\ \bibinfo
		{author} {\bibfnamefont {C.}~\bibnamefont {Hoffmann}},\ }\bibfield  {title}
	{\bibinfo {title} {Crystalplan: an experiment-planning tool for
			crystallography},\ }\href {https://doi.org/10.1107/S0021889811007102}
	{\bibfield  {journal} {\bibinfo  {journal} {Journal of Applied
				Crystallography}\ }\textbf {\bibinfo {volume} {44}},\ \bibinfo {pages} {418}
		(\bibinfo {year} {2011})}\BibitemShut {NoStop}%
	\bibitem [{\citenamefont {Rodríguez-Carvajal}(1993)}]{Fullprof}%
	\BibitemOpen
	\bibfield  {author} {\bibinfo {author} {\bibfnamefont {J.}~\bibnamefont
			{Rodríguez-Carvajal}},\ }\bibfield  {title} {\bibinfo {title} {Recent
			advances in magnetic structure determination by neutron powder diffraction},\
	}\href {https://doi.org/http://dx.doi.org/10.1016/0921-4526(93)90108-I}
	{\bibfield  {journal} {\bibinfo  {journal} {Physica B: Condensed Matter}\
		}\textbf {\bibinfo {volume} {192}},\ \bibinfo {pages} {55 } (\bibinfo {year}
		{1993})}\BibitemShut {NoStop}%
	\bibitem [{\citenamefont {Pet{\v{r}}{\'\i}{\v{c}}ek}\ \emph
		{et~al.}(2014)\citenamefont {Pet{\v{r}}{\'\i}{\v{c}}ek}, \citenamefont
		{Du{\v{s}}ek},\ and\ \citenamefont {Palatinus}}]{Petricek_2014}%
	\BibitemOpen
	\bibfield  {author} {\bibinfo {author} {\bibfnamefont {V.}~\bibnamefont
			{Pet{\v{r}}{\'\i}{\v{c}}ek}}, \bibinfo {author} {\bibfnamefont
			{M.}~\bibnamefont {Du{\v{s}}ek}},\ and\ \bibinfo {author} {\bibfnamefont
			{L.}~\bibnamefont {Palatinus}},\ }\bibfield  {title} {\bibinfo {title}
		{Crystallographic computing system {JANA2006}: General features},\ }\href
	{https://doi.org/doi:10.1515/zkri-2014-1737} {\bibfield  {journal} {\bibinfo
			{journal} {Zeitschrift für Kristallographie - Crystalline Materials}\
		}\textbf {\bibinfo {volume} {229}},\ \bibinfo {pages} {345} (\bibinfo {year}
		{2014})}\BibitemShut {NoStop}%
	\bibitem [{\citenamefont {Granroth}\ \emph {et~al.}(2006)\citenamefont
		{Granroth}, \citenamefont {Vandergriff},\ and\ \citenamefont
		{Nagler}}]{Granroth2006}%
	\BibitemOpen
	\bibfield  {author} {\bibinfo {author} {\bibfnamefont {G.~E.}\ \bibnamefont
			{Granroth}}, \bibinfo {author} {\bibfnamefont {D.~H.}\ \bibnamefont
			{Vandergriff}},\ and\ \bibinfo {author} {\bibfnamefont {S.~E.}\ \bibnamefont
			{Nagler}},\ }\bibfield  {title} {\bibinfo {title} {Sequoia: A fine resolution
			chopper spectrometer at the {SNS}},\ }\href
	{https://doi.org/10.1016/j.physb.2006.05.379} {\bibfield  {journal} {\bibinfo
			{journal} {Physica B: Condensed Matter}\ }\textbf {\bibinfo {volume}
			{385-86}},\ \bibinfo {pages} {1104} (\bibinfo {year} {2006})}\BibitemShut
	{NoStop}%
	\bibitem [{\citenamefont {Granroth}\ \emph {et~al.}(2010)\citenamefont
		{Granroth}, \citenamefont {Kolesnikov}, \citenamefont {Sherline},
		\citenamefont {Clancy}, \citenamefont {Ross}, \citenamefont {Ruff},
		\citenamefont {Gaulin},\ and\ \citenamefont {Nagler}}]{Granroth2010}%
	\BibitemOpen
	\bibfield  {author} {\bibinfo {author} {\bibfnamefont {G.~E.}\ \bibnamefont
			{Granroth}}, \bibinfo {author} {\bibfnamefont {A.~I.}\ \bibnamefont
			{Kolesnikov}}, \bibinfo {author} {\bibfnamefont {T.~E.}\ \bibnamefont
			{Sherline}}, \bibinfo {author} {\bibfnamefont {J.~P.}\ \bibnamefont
			{Clancy}}, \bibinfo {author} {\bibfnamefont {K.~A.}\ \bibnamefont {Ross}},
		\bibinfo {author} {\bibfnamefont {J.~P.~C.}\ \bibnamefont {Ruff}}, \bibinfo
		{author} {\bibfnamefont {B.~D.}\ \bibnamefont {Gaulin}},\ and\ \bibinfo
		{author} {\bibfnamefont {S.~E.}\ \bibnamefont {Nagler}},\ }\bibfield  {title}
	{\bibinfo {title} {Sequoia: A newly operating chopper spectrometer at the
			{SNS}},\ }\href {http://stacks.iop.org/1742-6596/251/i=1/a=012058} {\bibfield
		{journal} {\bibinfo  {journal} {Journal of Physics: Conference Series}\
		}\textbf {\bibinfo {volume} {251}},\ \bibinfo {pages} {012058} (\bibinfo
		{year} {2010})}\BibitemShut {NoStop}%
	\bibitem [{\citenamefont {Mason}\ \emph {et~al.}(2006)\citenamefont {Mason},
		\citenamefont {Abernathy}, \citenamefont {Anderson}, \citenamefont {Ankner},
		\citenamefont {Egami}, \citenamefont {Ehlers}, \citenamefont {Ekkebus},
		\citenamefont {Granroth}, \citenamefont {Hagen}, \citenamefont {Herwig},
		\citenamefont {Hodges}, \citenamefont {Hoffmann}, \citenamefont {Horak},
		\citenamefont {Horton}, \citenamefont {Klose}, \citenamefont {Larese},
		\citenamefont {Mesecar}, \citenamefont {Myles}, \citenamefont {Neuefeind},
		\citenamefont {Ohl}, \citenamefont {Tulk}, \citenamefont {Wang},\ and\
		\citenamefont {Zhao}}]{mason2006spallation}%
	\BibitemOpen
	\bibfield  {author} {\bibinfo {author} {\bibfnamefont {T.~E.}\ \bibnamefont
			{Mason}}, \bibinfo {author} {\bibfnamefont {D.}~\bibnamefont {Abernathy}},
		\bibinfo {author} {\bibfnamefont {I.}~\bibnamefont {Anderson}}, \bibinfo
		{author} {\bibfnamefont {J.}~\bibnamefont {Ankner}}, \bibinfo {author}
		{\bibfnamefont {T.}~\bibnamefont {Egami}}, \bibinfo {author} {\bibfnamefont
			{G.}~\bibnamefont {Ehlers}}, \bibinfo {author} {\bibfnamefont
			{A.}~\bibnamefont {Ekkebus}}, \bibinfo {author} {\bibfnamefont
			{G.}~\bibnamefont {Granroth}}, \bibinfo {author} {\bibfnamefont
			{M.}~\bibnamefont {Hagen}}, \bibinfo {author} {\bibfnamefont
			{K.}~\bibnamefont {Herwig}}, \bibinfo {author} {\bibfnamefont
			{J.}~\bibnamefont {Hodges}}, \bibinfo {author} {\bibfnamefont
			{C.}~\bibnamefont {Hoffmann}}, \bibinfo {author} {\bibfnamefont
			{C.}~\bibnamefont {Horak}}, \bibinfo {author} {\bibfnamefont
			{L.}~\bibnamefont {Horton}}, \bibinfo {author} {\bibfnamefont
			{F.}~\bibnamefont {Klose}}, \bibinfo {author} {\bibfnamefont
			{J.}~\bibnamefont {Larese}}, \bibinfo {author} {\bibfnamefont
			{A.}~\bibnamefont {Mesecar}}, \bibinfo {author} {\bibfnamefont
			{D.}~\bibnamefont {Myles}}, \bibinfo {author} {\bibfnamefont
			{J.}~\bibnamefont {Neuefeind}}, \bibinfo {author} {\bibfnamefont
			{M.}~\bibnamefont {Ohl}}, \bibinfo {author} {\bibfnamefont {C.}~\bibnamefont
			{Tulk}}, \bibinfo {author} {\bibfnamefont {X.-L.}\ \bibnamefont {Wang}},\
		and\ \bibinfo {author} {\bibfnamefont {J.}~\bibnamefont {Zhao}},\ }\bibfield
	{title} {\bibinfo {title} {The spallation neutron source in {Oak} {Ridge}: A
			powerful tool for materials research},\ }\href
	{https://doi.org/10.1016/j.physb.2006.05.281} {\bibfield  {journal} {\bibinfo
			{journal} {Physica B: Condensed Matter}\ }\textbf {\bibinfo {volume}
			{385}},\ \bibinfo {pages} {955} (\bibinfo {year} {2006})}\BibitemShut
	{NoStop}%
	\bibitem [{\citenamefont {Rule}\ \emph {et~al.}(2018)\citenamefont {Rule},
		\citenamefont {Mole},\ and\ \citenamefont {Yu}}]{rule2018glue}%
	\BibitemOpen
	\bibfield  {author} {\bibinfo {author} {\bibfnamefont {K.~C.}\ \bibnamefont
			{Rule}}, \bibinfo {author} {\bibfnamefont {R.~A.}\ \bibnamefont {Mole}},\
		and\ \bibinfo {author} {\bibfnamefont {D.}~\bibnamefont {Yu}},\ }\bibfield
	{title} {\bibinfo {title} {Which glue to choose? {A} neutron scattering study
			of various adhesive materials and their effect on background scattering},\
	}\href {https://doi.org/10.1107/S1600576718014930} {\bibfield  {journal}
		{\bibinfo  {journal} {Journal of Applied Crystallography}\ }\textbf {\bibinfo
			{volume} {51}},\ \bibinfo {pages} {1766} (\bibinfo {year}
		{2018})}\BibitemShut {NoStop}%
\end{thebibliography}

\begin{thebibliography}{13}%
	\makeatletter
	\providecommand \@ifxundefined [1]{%
		\@ifx{#1\undefined}
	}%
	\providecommand \@ifnum [1]{%
		\ifnum #1\expandafter \@firstoftwo
		\else \expandafter \@secondoftwo
		\fi
	}%
	\providecommand \@ifx [1]{%
		\ifx #1\expandafter \@firstoftwo
		\else \expandafter \@secondoftwo
		\fi
	}%
	\providecommand \natexlab [1]{#1}%
	\providecommand \enquote  [1]{``#1''}%
	\providecommand \bibnamefont  [1]{#1}%
	\providecommand \bibfnamefont [1]{#1}%
	\providecommand \citenamefont [1]{#1}%
	\providecommand \href@noop [0]{\@secondoftwo}%
	\providecommand \href [0]{\begingroup \@sanitize@url \@href}%
	\providecommand \@href[1]{\@@startlink{#1}\@@href}%
	\providecommand \@@href[1]{\endgroup#1\@@endlink}%
	\providecommand \@sanitize@url [0]{\catcode `\\12\catcode `\$12\catcode
		`\&12\catcode `\#12\catcode `\^12\catcode `\_12\catcode `\%12\relax}%
	\providecommand \@@startlink[1]{}%
	\providecommand \@@endlink[0]{}%
	\providecommand \url  [0]{\begingroup\@sanitize@url \@url }%
	\providecommand \@url [1]{\endgroup\@href {#1}{\urlprefix }}%
	\providecommand \urlprefix  [0]{URL }%
	\providecommand \Eprint [0]{\href }%
	\providecommand \doibase [0]{https://doi.org/}%
	\providecommand \selectlanguage [0]{\@gobble}%
	\providecommand \bibinfo  [0]{\@secondoftwo}%
	\providecommand \bibfield  [0]{\@secondoftwo}%
	\providecommand \translation [1]{[#1]}%
	\providecommand \BibitemOpen [0]{}%
	\providecommand \bibitemStop [0]{}%
	\providecommand \bibitemNoStop [0]{.\EOS\space}%
	\providecommand \EOS [0]{\spacefactor3000\relax}%
	\providecommand \BibitemShut  [1]{\csname bibitem#1\endcsname}%
	\let\auto@bib@innerbib\@empty
	%</preamble>
	\bibitem [{\citenamefont {Rodríguez-Carvajal}(1993)}]{Fullprof}%
	\BibitemOpen
	\bibfield  {author} {\bibinfo {author} {\bibfnamefont {J.}~\bibnamefont
			{Rodríguez-Carvajal}},\ }\bibfield  {title} {\bibinfo {title} {Recent
			advances in magnetic structure determination by neutron powder diffraction},\
	}\href {https://doi.org/http://dx.doi.org/10.1016/0921-4526(93)90108-I}
	{\bibfield  {journal} {\bibinfo  {journal} {Physica B: Condensed Matter}\
		}\textbf {\bibinfo {volume} {192}},\ \bibinfo {pages} {55 } (\bibinfo {year}
		{1993})}\BibitemShut {NoStop}%
	\bibitem [{\citenamefont {Bramwell}\ and\ \citenamefont
		{Holdsworth}(1993)}]{Bramwell_1993}%
	\BibitemOpen
	\bibfield  {author} {\bibinfo {author} {\bibfnamefont {S.~T.}\ \bibnamefont
			{Bramwell}}\ and\ \bibinfo {author} {\bibfnamefont {P.~C.~W.}\ \bibnamefont
			{Holdsworth}},\ }\bibfield  {title} {\bibinfo {title} {Magnetization and
			universal sub-critical behaviour in two-dimensional {XY} magnets},\ }\href
	{https://doi.org/10.1088/0953-8984/5/4/004} {\bibfield  {journal} {\bibinfo
			{journal} {Journal of Physics: Condensed Matter}\ }\textbf {\bibinfo {volume}
			{5}},\ \bibinfo {pages} {L53} (\bibinfo {year} {1993})}\BibitemShut {NoStop}%
	\bibitem [{\citenamefont {Kosterlitz}(1974)}]{Kosterlitz_1974}%
	\BibitemOpen
	\bibfield  {author} {\bibinfo {author} {\bibfnamefont {J.~M.}\ \bibnamefont
			{Kosterlitz}},\ }\bibfield  {title} {\bibinfo {title} {The critical
			properties of the two-dimensional xy model},\ }\href
	{https://doi.org/10.1088/0022-3719/7/6/005} {\bibfield  {journal} {\bibinfo
			{journal} {Journal of Physics C: Solid State Physics}\ }\textbf {\bibinfo
			{volume} {7}},\ \bibinfo {pages} {1046} (\bibinfo {year} {1974})}\BibitemShut
	{NoStop}%
	\bibitem [{\citenamefont {Cornelius}\ \emph {et~al.}(1986)\citenamefont
		{Cornelius}, \citenamefont {Day}, \citenamefont {Fyne}, \citenamefont
		{Hutchings},\ and\ \citenamefont {Walker}}]{Cornelius_1986}%
	\BibitemOpen
	\bibfield  {author} {\bibinfo {author} {\bibfnamefont {C.~A.}\ \bibnamefont
			{Cornelius}}, \bibinfo {author} {\bibfnamefont {P.}~\bibnamefont {Day}},
		\bibinfo {author} {\bibfnamefont {P.~J.}\ \bibnamefont {Fyne}}, \bibinfo
		{author} {\bibfnamefont {M.~T.}\ \bibnamefont {Hutchings}},\ and\ \bibinfo
		{author} {\bibfnamefont {P.~J.}\ \bibnamefont {Walker}},\ }\bibfield  {title}
	{\bibinfo {title} {Temperature and field dependence of the magnetisation of
			{Rb$_2$CrCl$_4$}: a two-dimensional easy-plane ionic ferromagnet},\ }\href
	{https://doi.org/10.1088/0022-3719/19/6/011} {\bibfield  {journal} {\bibinfo
			{journal} {Journal of Physics C: Solid State Physics}\ }\textbf {\bibinfo
			{volume} {19}},\ \bibinfo {pages} {909} (\bibinfo {year} {1986})}\BibitemShut
	{NoStop}%
	\bibitem [{\citenamefont {L{\'o}pez-Paz}\ \emph {et~al.}(2022)\citenamefont
		{L{\'o}pez-Paz}, \citenamefont {Guguchia}, \citenamefont {Pomjakushin},
		\citenamefont {Witteveen}, \citenamefont {Cervellino}, \citenamefont
		{Luetkens}, \citenamefont {Casati}, \citenamefont {Morpurgo},\ and\
		\citenamefont {von Rohr}}]{lopez2022dynamic}%
	\BibitemOpen
	\bibfield  {author} {\bibinfo {author} {\bibfnamefont {S.~A.}\ \bibnamefont
			{L{\'o}pez-Paz}}, \bibinfo {author} {\bibfnamefont {Z.}~\bibnamefont
			{Guguchia}}, \bibinfo {author} {\bibfnamefont {V.~Y.}\ \bibnamefont
			{Pomjakushin}}, \bibinfo {author} {\bibfnamefont {C.}~\bibnamefont
			{Witteveen}}, \bibinfo {author} {\bibfnamefont {A.}~\bibnamefont
			{Cervellino}}, \bibinfo {author} {\bibfnamefont {H.}~\bibnamefont
			{Luetkens}}, \bibinfo {author} {\bibfnamefont {N.}~\bibnamefont {Casati}},
		\bibinfo {author} {\bibfnamefont {A.~F.}\ \bibnamefont {Morpurgo}},\ and\
		\bibinfo {author} {\bibfnamefont {F.~O.}\ \bibnamefont {von Rohr}},\
	}\bibfield  {title} {\bibinfo {title} {Dynamic magnetic crossover at the
			origin of the hidden-order in van der waals antiferromagnet {CrSBr}},\ }\href
	{https://doi.org/10.48550/arXiv.2203.11785} {\bibfield  {journal} {\bibinfo
			{journal} {arXiv preprint arXiv:2203.11785}\ } (\bibinfo {year}
		{2022})}\BibitemShut {NoStop}%
	\bibitem [{\citenamefont {Telford}\ \emph {et~al.}(2020)\citenamefont
		{Telford}, \citenamefont {Dismukes}, \citenamefont {Lee}, \citenamefont
		{Cheng}, \citenamefont {Wieteska}, \citenamefont {Bartholomew}, \citenamefont
		{Chen}, \citenamefont {Xu}, \citenamefont {Pasupathy}, \citenamefont {Zhu}
		\emph {et~al.}}]{telford2020layered}%
	\BibitemOpen
	\bibfield  {author} {\bibinfo {author} {\bibfnamefont {E.~J.}\ \bibnamefont
			{Telford}}, \bibinfo {author} {\bibfnamefont {A.~H.}\ \bibnamefont
			{Dismukes}}, \bibinfo {author} {\bibfnamefont {K.}~\bibnamefont {Lee}},
		\bibinfo {author} {\bibfnamefont {M.}~\bibnamefont {Cheng}}, \bibinfo
		{author} {\bibfnamefont {A.}~\bibnamefont {Wieteska}}, \bibinfo {author}
		{\bibfnamefont {A.~K.}\ \bibnamefont {Bartholomew}}, \bibinfo {author}
		{\bibfnamefont {Y.-S.}\ \bibnamefont {Chen}}, \bibinfo {author}
		{\bibfnamefont {X.}~\bibnamefont {Xu}}, \bibinfo {author} {\bibfnamefont
			{A.~N.}\ \bibnamefont {Pasupathy}}, \bibinfo {author} {\bibfnamefont
			{X.}~\bibnamefont {Zhu}}, \emph {et~al.},\ }\bibfield  {title} {\bibinfo
		{title} {Layered antiferromagnetism induces large negative magnetoresistance
			in the van der {Waals} semiconductor {CrSBr}},\ }\href
	{https://doi.org/10.1002/adma.202003240} {\bibfield  {journal} {\bibinfo
			{journal} {Advanced Materials}\ }\textbf {\bibinfo {volume} {32}},\ \bibinfo
		{pages} {2003240} (\bibinfo {year} {2020})}\BibitemShut {NoStop}%
	\bibitem [{\citenamefont {Rule}\ \emph {et~al.}(2018)\citenamefont {Rule},
		\citenamefont {Mole},\ and\ \citenamefont {Yu}}]{rule2018glue}%
	\BibitemOpen
	\bibfield  {author} {\bibinfo {author} {\bibfnamefont {K.~C.}\ \bibnamefont
			{Rule}}, \bibinfo {author} {\bibfnamefont {R.~A.}\ \bibnamefont {Mole}},\
		and\ \bibinfo {author} {\bibfnamefont {D.}~\bibnamefont {Yu}},\ }\bibfield
	{title} {\bibinfo {title} {Which glue to choose? {A} neutron scattering study
			of various adhesive materials and their effect on background scattering},\
	}\href {https://doi.org/10.1107/S1600576718014930} {\bibfield  {journal}
		{\bibinfo  {journal} {Journal of Applied Crystallography}\ }\textbf {\bibinfo
			{volume} {51}},\ \bibinfo {pages} {1766} (\bibinfo {year}
		{2018})}\BibitemShut {NoStop}%
	\bibitem [{\citenamefont {Press}\ \emph {et~al.}(2007)\citenamefont {Press},
		\citenamefont {Teukolsky}, \citenamefont {Vetterling},\ and\ \citenamefont
		{Flannery}}]{NumericalRecipes}%
	\BibitemOpen
	\bibfield  {author} {\bibinfo {author} {\bibfnamefont {W.~H.}\ \bibnamefont
			{Press}}, \bibinfo {author} {\bibfnamefont {S.~A.}\ \bibnamefont
			{Teukolsky}}, \bibinfo {author} {\bibfnamefont {W.~T.}\ \bibnamefont
			{Vetterling}},\ and\ \bibinfo {author} {\bibfnamefont {B.~P.}\ \bibnamefont
			{Flannery}},\ }\href@noop {} {\emph {\bibinfo {title} {Numerical recipes 3rd
				edition: The art of scientific computing}}}\ (\bibinfo  {publisher}
	{Cambridge university press},\ \bibinfo {year} {2007})\BibitemShut {NoStop}%
	\bibitem [{\citenamefont {Scheie}\ \emph {et~al.}(2022)\citenamefont {Scheie},
		\citenamefont {Laurell}, \citenamefont {McClarty}, \citenamefont {Granroth},
		\citenamefont {Stone}, \citenamefont {Moessner},\ and\ \citenamefont
		{Nagler}}]{scheie2022spin}%
	\BibitemOpen
	\bibfield  {author} {\bibinfo {author} {\bibfnamefont {A.}~\bibnamefont
			{Scheie}}, \bibinfo {author} {\bibfnamefont {P.}~\bibnamefont {Laurell}},
		\bibinfo {author} {\bibfnamefont {P.~A.}\ \bibnamefont {McClarty}}, \bibinfo
		{author} {\bibfnamefont {G.~E.}\ \bibnamefont {Granroth}}, \bibinfo {author}
		{\bibfnamefont {M.~B.}\ \bibnamefont {Stone}}, \bibinfo {author}
		{\bibfnamefont {R.}~\bibnamefont {Moessner}},\ and\ \bibinfo {author}
		{\bibfnamefont {S.~E.}\ \bibnamefont {Nagler}},\ }\bibfield  {title}
	{\bibinfo {title} {Spin-exchange hamiltonian and topological degeneracies in
			elemental gadolinium},\ }\href {https://doi.org/10.1103/PhysRevB.105.104402}
	{\bibfield  {journal} {\bibinfo  {journal} {Phys. Rev. B}\ }\textbf {\bibinfo
			{volume} {105}},\ \bibinfo {pages} {104402} (\bibinfo {year}
		{2022})}\BibitemShut {NoStop}%
	\bibitem [{\citenamefont {Stoner}\ and\ \citenamefont
		{Wohlfarth}(1948)}]{stoner1948mechanism}%
	\BibitemOpen
	\bibfield  {author} {\bibinfo {author} {\bibfnamefont {E.~C.}\ \bibnamefont
			{Stoner}}\ and\ \bibinfo {author} {\bibfnamefont {E.}~\bibnamefont
			{Wohlfarth}},\ }\bibfield  {title} {\bibinfo {title} {A mechanism of magnetic
			hysteresis in heterogeneous alloys},\ }\href
	{https://doi.org/10.1098/rsta.1948.0007} {\bibfield  {journal} {\bibinfo
			{journal} {Philosophical Transactions of the Royal Society of London. Series
				A, Mathematical and Physical Sciences}\ }\textbf {\bibinfo {volume} {240}},\
		\bibinfo {pages} {599} (\bibinfo {year} {1948})}\BibitemShut {NoStop}%
	\bibitem [{\citenamefont {Zhdanova}\ \emph {et~al.}(2011)\citenamefont
		{Zhdanova}, \citenamefont {Lyakhova},\ and\ \citenamefont
		{Pastushenkov}}]{zhdanova2011magnetocrystalline}%
	\BibitemOpen
	\bibfield  {author} {\bibinfo {author} {\bibfnamefont {O.}~\bibnamefont
			{Zhdanova}}, \bibinfo {author} {\bibfnamefont {M.}~\bibnamefont {Lyakhova}},\
		and\ \bibinfo {author} {\bibfnamefont {Y.~G.}\ \bibnamefont {Pastushenkov}},\
	}\bibfield  {title} {\bibinfo {title} {Magnetocrystalline anisotropy,
			magnetization curves, and domain structure of {FeB} single crystals},\ }\href
	{https://doi.org/10.1134/S0031918X11030306} {\bibfield  {journal} {\bibinfo
			{journal} {The Physics of Metals and Metallography}\ }\textbf {\bibinfo
			{volume} {112}},\ \bibinfo {pages} {224} (\bibinfo {year}
		{2011})}\BibitemShut {NoStop}%
	\bibitem [{\citenamefont {Zhdanova}\ \emph {et~al.}(2013)\citenamefont
		{Zhdanova}, \citenamefont {Lyakhova},\ and\ \citenamefont
		{Pastushenkov}}]{Zhdanova2013}%
	\BibitemOpen
	\bibfield  {author} {\bibinfo {author} {\bibfnamefont {O.~V.}\ \bibnamefont
			{Zhdanova}}, \bibinfo {author} {\bibfnamefont {M.~B.}\ \bibnamefont
			{Lyakhova}},\ and\ \bibinfo {author} {\bibfnamefont {Y.~G.}\ \bibnamefont
			{Pastushenkov}},\ }\bibfield  {title} {\bibinfo {title} {Magnetic properties
			and domain structure of {FeB} single crystals},\ }\href
	{https://doi.org/10.1007/s11041-013-9581-0} {\bibfield  {journal} {\bibinfo
			{journal} {Metal Science and Heat Treatment}\ }\textbf {\bibinfo {volume}
			{55}},\ \bibinfo {pages} {68} (\bibinfo {year} {2013})}\BibitemShut {NoStop}%
	\bibitem [{\citenamefont {Bae}\ \emph {et~al.}(2022)\citenamefont {Bae},
		\citenamefont {Wang}, \citenamefont {Xu}, \citenamefont {Chica},
		\citenamefont {Diederich}, \citenamefont {Cenker}, \citenamefont {Ziebel},
		\citenamefont {Bai}, \citenamefont {Ren}, \citenamefont {Dean} \emph
		{et~al.}}]{bae2022exciton}%
	\BibitemOpen
	\bibfield  {author} {\bibinfo {author} {\bibfnamefont {Y.~J.}\ \bibnamefont
			{Bae}}, \bibinfo {author} {\bibfnamefont {J.}~\bibnamefont {Wang}}, \bibinfo
		{author} {\bibfnamefont {J.}~\bibnamefont {Xu}}, \bibinfo {author}
		{\bibfnamefont {D.~G.}\ \bibnamefont {Chica}}, \bibinfo {author}
		{\bibfnamefont {G.~M.}\ \bibnamefont {Diederich}}, \bibinfo {author}
		{\bibfnamefont {J.}~\bibnamefont {Cenker}}, \bibinfo {author} {\bibfnamefont
			{M.~E.}\ \bibnamefont {Ziebel}}, \bibinfo {author} {\bibfnamefont
			{Y.}~\bibnamefont {Bai}}, \bibinfo {author} {\bibfnamefont {H.}~\bibnamefont
			{Ren}}, \bibinfo {author} {\bibfnamefont {C.~R.}\ \bibnamefont {Dean}}, \emph
		{et~al.},\ }\bibfield  {title} {\bibinfo {title} {Exciton-coupled coherent
			antiferromagnetic magnons in a 2d semiconductor},\ }\href
	{https://doi.org/10.48550/arXiv.2201.13197} {\bibfield  {journal} {\bibinfo
			{journal} {arXiv preprint arXiv:2201.13197}\ } (\bibinfo {year}
		{2022})}\BibitemShut {NoStop}%
	
\end{thebibliography}
\end{document}